\documentclass[12pt]{article}

\usepackage{geometry}
\usepackage{amsfonts}
\usepackage{stmaryrd}
\usepackage{amssymb}
\usepackage{euscript}
\usepackage{amsthm}
\usepackage{latexsym,amsmath,amscd}
\usepackage{mathrsfs}
\usepackage{graphicx}
\usepackage{color}
\usepackage{dsfont}
\usepackage{enumerate}
\usepackage{booktabs}

\usepackage{caption}
\usepackage{subcaption}

\newtheorem{theorem}{Theorem}[section]

\newtheorem{lem}{Lemma}[section]
\newtheorem{pro}{Proposition}[section]
\newtheorem{cor}{Corollary}[section]
\newtheorem{rem}{Remark}[section]
\newtheorem{rems}{Remarks}[section]
\newtheorem{ex}{Example}[section]
\newtheorem{defi}{Definition}[section]
\newtheorem{hyp}{Assumption}[section]

\newcommand{\bt}{\begin{theorem}}
\newcommand{\et}{\end{theorem}}
\newcommand{\bl}{\begin{lem}}
\newcommand{\el}{\end{lem}}
\newcommand{\bp}{\begin{pro}}
\newcommand{\ep}{\end{pro}}
\newcommand{\bcor}{\begin{cor}}
\newcommand{\ecor}{\end{cor}}
\newcommand{\lab }{\label }
\newcommand{\bd}{\begin{defi} \rm }
\newcommand{\ed}{\end{defi}}
\newcommand{\brem }{\begin{rem} \rm }
\newcommand{\erem }{\end{rem}}
\newcommand{\brems }{\begin{rems} \rm }
\newcommand{\erems }{\end{rems}}
\newcommand{\bhyp }{\begin{hyp} \rm }
\newcommand{\ehyp }{\end{hyp}}
\newcommand{\bex}{\begin{ex} \rm }
\newcommand{\eex}{\end{ex}}

\newcommand{\ssc}{\subsection}

\newcommand{\be}{\begin{equation}}
\newcommand{\ee}{\end{equation}}
\newcommand{\bde}{\begin{displaymath}}
\newcommand{\ede}{\end{displaymath}}
\newcommand{\beq}{\begin{eqnarray*}}
\newcommand{\eeq}{\end{eqnarray*}}
\newcommand{\beqa}{\begin{eqnarray}}
\newcommand{\eeqa}{\end{eqnarray}}
\newcommand{\bel }{\left\{\begin{array}{ll}}
\newcommand{\eel}{\cr \end{array} \right.}

\newcommand{\rosso}[1]{{\color{red} #1}}
\newcommand{\verde}[1]{{\color{blue} #1}}

\def\bhP{\widehat P}

\def\CVA{{\rm CVA}}

\def\FCVA{{\rm FCVA}}
\def\FRCVA{{\rm FRCVA}}
\def\FRA{{\rm FRA}}

\def\*BFCVA{{\rm (B)FCVA}}
\def\FVA{{\rm FVA}}
\def\WCFS{{\rm WACC}}
\def\WCFS{{\rm WCFS}}
\def\WCIS{{\rm WCIS}}

\def\EE{\mathbb{E}_{\mathbb{Q}}}

\def\EPP{\mathbb{E}_{\mathbb{P}}}

\def\FG{{\mathbb{G}}}

\def\P{\mathbb{P}}
\def\Q{\mathbb{Q}}

\def\G{{\cal{G}}}

\def\I{\mathds{1}}

\usepackage{eso-pic}

\title{\vspace{-1cm}{\Large \bf An initial approach to Risk Management \\ of Funding Costs\thanks{This paper represents the authors opinion and is in no way representative of the views of the institutions the authors work for or are associated with. The authors are grateful to Andrea Pallavicini for helpful feedback on a first draft. Corresponding author: {\tt c.durand12@imperial.ac.uk}}\vskip 0 pt}}
\author{{\normalsize Damiano Brigo} \\ 
{\normalsize Department of Mathematics}\\
{\normalsize Imperial College London} \\ \and 
{\normalsize Cyril Durand}\\
{\normalsize Department of Mathematics}\\
{\normalsize Imperial College London}\\
}

\date{\small{First version: 1 Dec 2013; This version: 7 Oct 2014}}

\begin{document}
\maketitle

\vspace{-0.5cm}

\begin{abstract}
In this note we sketch an initial tentative approach to funding costs analysis and management for contracts with bilateral counterparty risk in a simplified setting.
We depart from the existing literature by analyzing the issue of funding costs and benefits under the assumption that the associated risks cannot be hedged properly. 
We also model the treasury funding spread by means of a stochastic Weighted Cost of Funding Spread ($\WCFS$) which helps 
describing more realistic financing policies of a financial institution.
%and discuss right-or-wrong way funding risk as well as systemic funding risk.
% and analyzing the impact of the suitable modelling of margin agreements. 
% in the presence of right-or-wrong way risk, first-to-default and systemic risks.
We elaborate on some limitations in replication-based Funding / Credit Valuation Adjustments we worked on ourselves in the past, namely CVA, DVA, FVA and related quantities as generally discussed in the industry.  We advocate as a different possibility, when replication is not possible, the analysis of the funding profit and loss distribution 
%study funding loss model risk 
and explain how long term funding spreads, wrong way risk and systemic 
risk are generally overlooked in most of the current literature on risk measurement of funding costs.
As a matter of initial illustration, we discuss in detail the funding management of interest rate swaps with bilateral counterparty risk in the simplified setup of our framework through numerical examples and via a few simplified assumptions.

\end{abstract}

\vskip 10 pt \hskip 10 pt {\small {\bf Key words.}  Credit valuation adjustment, CVA, 
Funding valuation adjustment, FVA,  Funding risk adjustment, FRA, Funding risk credit 
valuation adjustment, FRCVA, Wrong way funding risk, 
Systemic funding risk, Interest rate swap, 
Weighted Cost of Funding Spread, \WCFS, Term structure of funding costs, Funding loss distribution}

\bigskip 

{\bf AMS} Classification Code: 62P5. {\bf JEL} Classification Code: C60

%\tableofcontents

\bigskip 

%%%%%%%%%%%%%%%%%%%%%%%%%%%%%%%%%%%%%%%%%%%%%%%%%%%%%%%%%%%%%%%%%%%%%%%%%%%%%%%%%%%%%%
%%%%%%%%%%%%%%%%%%%%%%%%%%%%%%%%%%%%%%%%%%%%%%%%%%%%%%%%%%%%%%%%%%%%%%%%%%%%%%%%%%%%%%
\section{Introduction} \label{sect1}
%%%%%%%%%%%%%%%%%%%%%%%%%%%%%%%%%%%%%%%%%%%%%%%%%%%%%%%%%%%%%%%%%%%%%%%%%%%%%%%%%%%%%%
%%%%%%%%%%%%%%%%%%%%%%%%%%%%%%%%%%%%%%%%%%%%%%%%%%%%%%%%%%%%%%%%%%%%%%%%%%%%%%%%%%%%%%

What is of practical concern to trading desks nowadays, as far as the risk management of funding costs and benefits is concerned, 
is essentially to build-up relevant decision tools that allow for  assessing and possibly hedging the risk their future carry costs bear on their profitability as well as gauging the costs and benefits of alternative contract specifications. 
These tools can be constructed by resorting to simulation engines, typically of Monte-Carlo type, that are required to account for the diffusion of the market risk drivers, the future movements of discount factors,  the evolution of the contract (or portfolio of contracts) marked-to-market value(s), the evolution of the credit qualities of the obligors, possibly until default, the contract (or portfolio of contracts) loss upon counterparty default, but also for the exchange of collaterals, the stochasticity of the funding spreads, the alternative hedging strategies, right-or-wrong way risk and systemic risk.

In this article we study a number of features that need to be considered when building-up these tools. In particular we examine in a specified setting the full distribution of funding losses, as well as right-or-wrong way risk, systemic risk, funding spread modelling,
and the asymmetry of contract profitability implied by the funding loss.

The organisation of the paper is as follows.

In section \ref{sect1}, we start by recalling basic definitions for bilateral counterparty credit risk and explain how this paper relates to the earlier literature, including our own, on credit and funding risk valuation.
We then introduce the concepts of  funding loss distribution and Funding Risk Assessment (FRA). We finally present the Weighted Cost of Funding Spread ($\WCFS$). 

We proceed in section \ref{sect2} by illustrating our approach through the examination of the credit valuation adjustment and the funding risk measurement for interest rate swaps, and providing numerical examples. We conclude with Section \ref{seccon} by listing the unaddressed problems and discussing further research.

%%%%%%%%%%%%%%%%%%%%%%%%%%%%%%%%%%%%%%%%%%%%%%%%%%%%%%%%%%%%%%%%%%%%%%%%%%%%%%%%%%%%%%
%%%%%%%%%%%%%%%%%%%%%%%%%%%%%%%%%%%%%%%%%%%%%%%%%%%%%%%%%%%%%%%%%%%%%%%%%%%%%%%%%%%%%%
\section{Funding Loss Distribution} \label{sect1}
%%%%%%%%%%%%%%%%%%%%%%%%%%%%%%%%%%%%%%%%%%%%%%%%%%%%%%%%%%%%%%%%%%%%%%%%%%%%%%%%%%%%%%
%%%%%%%%%%%%%%%%%%%%%%%%%%%%%%%%%%%%%%%%%%%%%%%%%%%%%%%%%%%%%%%%%%%%%%%%%%%%%%%%%%%%%%

%%%%%%%%%%%%%%%%%%%%%%%%%%%%%%%%%%%%%%%%%%%%%%%%%%%%%%%%%%%%%%%%%%%%%
%\ssc{Preliminary Comment on Funding}  
%%%%%%%%%%%%%%%%%%%%%%%%%%%%%%%%%%%%%%%%%%%%%%%%%%%%%%%%%%%%%%%%%%%%%

\subsection{Funding Costs and Benefits: Risk Management or Valuation?} 

The examination of credit and funding adjustments and costs and of their implications on pricing has become a sensitive subject in the realm of finance since the realization that the assumption under which financial institutions can trade without default risk and fund their trading activities at the 'risk free rate' is not realistic post financial crisis. Already in Brigo in Masetti (2005) the Credit Valuation Adjustment for interest rate swap portfolios and equity return swaps is introduced in detail in derivatives valuation. 
Most banks nowadays price their collateralized deals by means of dedicated discount curves (Fuji et al \cite{FST}, Piterbarg \cite{P}), and aim at evaluating the possibly significant impact of funding costs, especially on their uncollateralized transactions. As a consequence, heightened attention has been given to proposing unified frameworks for computing adjustments to the risk free price due to both funding and credit effects, resulting first in separate valuation adjustments such as Credit Valuation Adjustments (CVA) and Funding Valuation Adjustments (FVA), or in non-separable adjustments (FCVA) (Cr\'epey \cite{C}, Pallavicini et al. \cite{PPB}, Burgard and Kjaer \cite{BK}, Kenyon et al. \cite{KS}). This approach utilizes classical derivatives valuation and a ``replication approach", even if invariance results in \cite{PPB} show that the final valuation master equations depend only on observable market rates and not on the risk neutral theoretical rate of the risk neutral measure, see also Elouerkhaoui (2014). 

While in many situations the above "replication approach" can be considered adequate and a good approximation, in this paper we play devil's advocate and we propose an alternative approach in case the financial institution cannot dynamically
replicate funding costs themselves. In this sense we depart from the ``replication approach" based on traditional derivatives valuation and leading to the different XVA's above. 

%Our main focus is on elaborating on some important aspects of funding in view of the construction of efficient funding risk management tools. 
%The valuation and hedging of funding is a complex issue which requires not only the modelling, pricing and netting of a large portfolio of financial contracts (see Brigo \cite{B}, Cesari %\cite{CACFLM}, Durand and Rutkowski \cite{DR}) but also the analysis of a financial institution' policies in terms of funding \& liquidity, internal accounting rules, CSA contracts %management etc., as well as the examination of its trading books and technological platforms so as to design and implement a winning 'Corporate Funding and $\CVA$  strategy' (see %Durand \cite{D1} for a general description of an applied methodology - Ask Damiano if OK). 

%%%%%%%%%%%%%%%%%%%%%%%%%%%%%%%%%%%%%%%%%%%%%%%%%%%%%%%%%%%%%%%%%%%%%
\ssc{Basic Definitions: Credit risk, collateral and CVA}  \label{sect2.1}
%%%%%%%%%%%%%%%%%%%%%%%%%%%%%%%%%%%%%%%%%%%%%%%%%%%%%%%%%%%%%%%%%%%%%

We start by recalling basic definitions regarding bilateral counterparty risk. We partly follow Durand and Rutkowski (2013). 
The obligors of the contract are refered to as {\it the investor}
and {\it the counterparty}. 
Let $\tau^1 $ and $\tau^2 $ stand for the random times
of default of the investor and the counterparty. More generally, quantities indexed by $1$ will refer to the investor and by $2$ to the counterparty.  
They are defined on an underlying probability space $(\Omega , {\cal G},\Q)$,
where $\Q$ is the risk-neutral probability measure.
We also denote by $\EE$ the expectation under $\Q$ and by $\G_t$ the $\sigma$-field of all events observed
by time $t$. It is assumed throughout that $\tau^1 $ and $\tau^2 $ are stopping times with respect to the filtration
$\FG  = (\G_t)_{t \geq 0}$. 

\bd  \lab{lgd1}
The investor's {\it loss given default} $L^{1}_t$  represents the loss incurred
by the %non-defaulting 
investor in the event of the counterparty default at time $t$.
It is equal to the difference between the replacement cost and the
settlement value of the contract at the moment of default, that is, $L^{1}_t := P_t^{*} - S_t$.
\ed

In practice, and even tough this approach is not fully in agreement with ISDA recommandations (see for instance \cite{BMRisk} and  \cite{DR} for a detailed analysis of these
discrepancies),
%the settlement value is computed as a function of the "counterparty risk free contract" $P_t$ whilst 
the equivalent contract is considered equal to the risk free contract ($P^*_t = P_t$). 

%\bd \lab{defwr}
%The {\it loss process} $\mathcal{L}^{1}$ represents the investor's loss given default process, specifically,
%$\mathcal{L}^{i}_t := \I_{\{ t \ge  \tau^2 \}} \, L^{1}_{\tau^2}$.
%\ed

We now introduce the following

\bd
The loss processes are defined as ${\cal L}^1_t = \I_{\{ t \ge  \tau^2 \}} \, L^{1}_{\tau^2}$ and 
${\cal L}^2_t = \I_{\{ t \ge  \tau^1 \}} \, L^{2}_{\tau^1}$.
\ed

The next definition introduces wrong and right way counterparty credit risks.
Counterparty credit risk is wrong (right, resp.) way if the exposure tends to increase (decrease, resp.) when
the counterparty credit quality worsens.  

\bd \label{def2.15c}
The {\it investor's right} ({\it wrong}, resp.) {\it way counterparty credit risk} is the positive (negative, resp.) dependence between the counterparty's
creditworthiness  
%$c^2$ 
and the {\it positive loss} process $\mathcal{L}^{1,+}_t := \big[ \mathcal{L}^{1}_t \big]^+ = \I_{\{ t \ge  \tau^2 \}} \, [L^{1}_{\tau^2}]^+ $.
\ed

\bd \lab{sv1}
In the absence of collateralization, the contract's {\it settlement value} $S_t$  in case of the
counterparty's default at time $t$ is given by the stochastic process $S_t :=
R^2_t (P_t)^+ - (P_t)^-$ where $R^2_t \in [0,1]$ is the counterparty's recovery rate.
\ed

\bp  \lab{lgd1}
In the absence of collateralization, the investor's {\it loss given default} $L^{1}_t$ 
is given by 
\beq 
L^{1}_t := P_t - S_t = P_t - \big( R^2_t (P_t)^+ - (P_t)^-  \big) = \big(  1 - R^2_t \big) (P_t)^+.
\eeq
\ep

We proceed by adding collateralization in our setting (as seen by the investor).

\bd
The collateral process  $C_t$ represents the exchange of collaterals as specified in a Credit Support Annex.
%:
%\bde
%C_t = C^+_t - C^{-}_t = \I_{\{ C_t \geq 0\}} \, C^2_t  - \I_{\{ C_t < 0\}} \, C^1_t ,
%\ede
%where $C^1_t$ ($C^2_t$, resp.) denotes the market value of the basket of collaterals posted at time $t$ by the
%investor (the counterparty, resp.).
\ed

%The next result provides the loss given default, as seen by the investor, in the presence of a collateral agreement.

\bl
On the event  $ \{ \tau^2 = t \le T, \tau^1 > \tau^2 \} $ of counterparty's default,
the loss given default for the investor (in the absence of segregation and initial margin) equals
\beq %\lab{6A}
L^1_t = (1- R^2_t ) \, \Big(  \I_{\{ P_t \geq 0\}} \, (P_t - C^+_t )^+ + \I_{\{ P_t  < 0\}} \, (P_t + C^-_t )^+\Big).
\eeq
where $C^-_t$ ($C^+_t$, resp.) denotes the market value of the basket of collaterals posted at time $t$ by the
investor (the counterparty, resp.).
\el

We are now in the position to introduce the definition of the {\it credit value adjustment}.

\bd \lab{cvai}
The {\it credit value adjustment} equals the difference between the risk free contract $P_t$
and the contract with counterparty credit risk $\bhP_t$:
\beq\CVA_t = P_t - \bhP_t.
\eeq
\ed 

\bp \lab{PC9}
The price $\bhP_t$ of a contract with bilateral counterparty risk is equal to the difference
of the risk-free price $P_t$ of the contract and the price of the protection leg of a contingent CDS
written on the investor's loss $\mathcal{L}^{1}$ due to the counterparty's default before the contract termination $T$,
augmented by the price of the protection leg of a contingent CDS
written on the counterparty's loss $\mathcal{L}^{2}$ due to the investor's default before the contract termination $T$.
This means that the following equality holds, for every $t \in [0,T]$ on the event $\{ t < \tau^1 \wedge \tau^2 \}$,
\be \nonumber
\CVA_t = P_t + P_t(\mathcal{L}^{1}) - P_t(\mathcal{L}^{2}).
\ee
\ep

\ssc{Specifying the Funding Loss Distribution} 

We adopt a simplifying scheme where the funding of a financial institution's activities is managed by the Treasury and the Finance departments by means of  a combination of short and long term borrowings. These are entered following the business inflows and outflows forecasts, including the exchange of collaterals. Finding a good hedge for these costs, at portfolio or stand alone product level, is a difficult task in practice essentially because a bank is not allowed to  hedge against movements in its own credit quality by resorting to credit default swaps on its own name. Hence, typically, only partial hedging can be achieved by using proxies such as indices or bonds.
It follows that, when valuing a financial instrument,  the inclusion of funding costs based on risk neutral valuation is debatable. 

In this paper we take the view of a financial institution that cannot dynamically replicate its funding costs and 
we examine the funding costs from a risk measurement standpoint rather than from a valuation point of view. 
We borrow the typical risk management terminology and introduce the 'funding loss' distribution of a contract or portfolio under  the physical or real world probability measure $\P$.
%Let us mention here that usually CVA is computed under the risk neutral measure, so that it can be necessary
%to apply a change of measure so as to ensure that CVA and funding are computed consistently when using a common engine. 

Our approach entails that we see funding as an indirect cost specific to each obligor similarly to, for instance, operational costs. 
We do not include funding into the 'replication price' of a contract (as in \cite{PPB}, \cite{C}) but instead consider funding risk analysis as a decision making component to assess the contract profitability. Notice that this approach offers an advantage in terms of computation since we do not have to deal with the complexity that follows from the integration of funding costs into nonlinear valuation (see \cite{C},\cite{PPB},\cite{BPCCP},\cite{NVA}).
Also, our methodology fits the perspective of a financial institution that is planning to manage funding risk through a global 'macro-hedging strategy', meaning that the financial institution manages globally the funding costs of the whole portfolio of financial instruments.

We now proceed by specifying what we mean by funding costs and risks. To this end, let us consider the processes that can 
potentially require cash lending or borrowing from the trading desks. 

Firstly, the investor may decide to hedge its contract obligations as well as associated CVA, i.e dynamically construct a self financing replicating portfolio for both the risk free contract and the CVA.  Secondly, the investor may be required to implement full or partial collateralization of the contract. Thirdly, the investor might need to fund a residual loss (or cash in a benefit) in the event of counterparty default. By residual loss due to default we mean for instance the loss that is not covered by the 
replicating CVA portfolio due to discrepancies between the equivalent contract, generally supposed to be risk free in the hedging calculations,  and the actual replacement contract upon default
(see \cite{BMRisk} and \cite{DR} for a discussion on the impact of varied settlement conventions upon default). We may also include other potential losses such as those incurred in when selling collateral under a period of liquidity constraints.
Each of these components, which we will refer to as the hedging process $H_t$, the collateralization process $C_t$ (introduced in the previous section) and the loss process
$L_t$ (not to be mistaken for $\mathcal{L}_t$),  potentially requires cash borrowing or lending, and hence an associated {\it funding cost}, leading to a cumulative funding loss. 

In the following we introduce the {\it funding requirement process}, as well as the {\it funding process},
before defining formally the funding loss.

\bd \lab{defwrf}
The {\it funding requirement process}  $f_t^{1}$  of a portfolio for the investor represents, at any time $t$, the investor's 
positive or negative funding requirements when financing or investing the cash flows given or received during the hedging process $H_t$, the collateralization  process $C_t$ and the loss process $L_t$. We will see later precise quantitative definitions for the process $f_t^{1}$. As an initial example, for an uncollateralized contract we may have
\[ f_t^{1} =    \big( -\bhP_t \big)^- -  \big( -\bhP_t \big)^+ \  . \]

\ed

%For example, in the case of a risk-free and fully collateralized financial instrument with hedging instruments traded in %swapped form, 
 %the funding requirement at any time t represents the collateral that needs to be posted (or is received), 
%i.e $f_t = C_t$.  
%Notice here that it implies that the funding requirement can be negative. This is the case, in the aforementioned example,
%when a collateral is received. 

\bd \lab{defwrf}
The {\it funding process} $\mathcal{F}_t^{1}$ of a portfolio for the investor represents, at any time $t$, the investor's funding cost or benefit 
when financing or investing the funding requirement at time $t$. We will see later precise quantitative definitions for the process $\mathcal{ F}_t^{1}$. As an initial example, for an uncollateralized contract we may have
\[ \mathcal{ F}_t^{1} =   \phi^{+}_t \, \big( -\bhP_t \big)^- - \phi^{-}_t \, \big( -\bhP_t \big)^+ \  , \]
where $\phi^+$ is the interest rate we pay on borrowed amounts and $\phi^-$ is the interest rate we receive on lent amounts. 
\ed

%To rebound on our previous example, the funding process at some time $t$ where the investor needs to borrow collateral is %characterized by a funding cost equal to the interest paid to the lender on the collateral amount for the associated borrowing %period. In continuous time, denoting by $\phi^1_t$ the funding spread of the investor prevailing at time $t$, the funding %process reads $\mathcal{F}_t^{1} = \phi^1_t f_t dt = \phi^1_t C_t dt$.

\bd
The {\it funding loss}  $\Phi^{1}$ of a portfolio with maturity $T$ for the investor is the cumulated discounted funding costs or benefits 
until contract termination or obligor default, that is
\beq
 \Phi^{1} = \int_0^{\tau^1 \wedge \tau^2 \wedge T} D(0,t) \,  \mathcal{F}^{1}_t dt.
\eeq
\ed
where $D(0,t)$ denotes the discount factor for time $t$.

\brem {\bf Which discount?}
It is important to point out here that we are discounting using a theoretical discount factor that has no clear direct interpretation, although 
considering funding costs as indirect costs naturally leads to discounting at some risk free rate, typically OIS/overnight.
More generally, we might as well omit discounting in a risk measurement approach. However, we feel that since we are addressing funding costs, discounting should be included in the loss being measured. Even so, despite the above argument for OIS, it is not clear at which rate we should really discount.  Adopt OIS as a proxy of the risk free rate would be a compromise, since the real discounting rate we will apply might depend on the actual funding policy itself, leading then to a recursive type equation. Even in the risk neutral valuation approach the recursion was found to be there, but in that context the theoretical risk free rate would disappear in an exact way from the equations, see for example \cite{PPB,BPCCP,MBP}, due to an invariance result. Here, for the time being, we will assume we are discounting at overnight, but this is an aspect of the analysis that needs further discussion. 
\erem

Note that the funding loss may be either positive or negative, in which latter
case it represents a gain. 
We are now in the position to introduce the {\it funding distribution} of a contract.

\bd
The {\it  $\P$-funding distribution} (or the {\it funding distribution}  when no confusion arises) of a contract for the investor is the distribution of the funding loss $ \Phi^{1} $  under the real measure $\P$.
\ed

In this article we argue that the funding distribution, that is usually not discussed much in the literature, is a key tool as far as funding risk management is concerned. In particular, the risk neutral or even $\P$ average of the funding cost cash flows typically do not provide sufficient information on the funding risk.

Let us recall here that the average funding costs, as typically computed when pricing a deal with funding and credit risk under the risk neutral measure, is related to what is commonly referred to as the 'funding valuation adjustment'. 

\bd
A simplified version of the {\it Funding Valuation Adjustment} of a contract for the investor $\FVA^{1}$ is the mean of the funding distribution of the contract when computed under the risk neutral measure $\Q$:
\beq
\FVA^{1} = \,  \EE  \big[ \Phi^{1} \big]  .
\eeq 
\ed

In our framework it is natural to introduce a risk measurement counterparty to the valuation one above. This measurement should be used to assess contract profitability and risk rather than a precise contract valuation to be quoted to a client or an internal desk. We call this quantity the  {\it Funding Risk Assessment}

\bd
The {\it Funding Risk Assessment } $\FRA^{1}$  of a contract for the investor is a risk statistics extracted from the funding loss.  It can be a quantile based on a given confidence level, a tail expectation beyond that quantile, etc. 
\ed 

For instance, if the distribution loss exhibits a very low variance and kurtosis, then
the FRA could be chosen as the mean of the loss distribution. However, in case of a large variance or kurtosis, the financial institution may decide to resort to a  conservative quantile, say 95$\%$, in order to almost fully cover its risk. More generally, the characterization of the FRA would depend not only on the funding risk profiles at single name and portfolio levels, but also on the general policy of the financial institution regarding funding risk management, and potential regulatory constraints. We can now formulate the following proposition, which essentially comes from the fact 
that the perfect replication of hedging costs is typically not possible in our setting.

\bp
The Funding Risk Assessment  is typically not equal to either the funding valuation adjustment
\beq
\FRA^{1} \neq \FVA^{1}
\eeq
or the risk neutral expectation of the funding distribution
\beq
\FRA^{1}  \neq \EE  \big[  \Phi^{1} \big] .
\eeq 
\ep

We finally introduce the concept of  {\it Funding Risk Credit Valuation Adjustment}, which also departs from the 
the Funding Credit Valuation Adjustment $\FCVA$ as typically found in the literature.

\bd
The {\it Funding Risk Credit Valuation Adjustment} $\FRCVA^{1}$ of a contract for the investor is the adjustment to the profitability of the risk-free contract due to bilateral counterparty credit risk and the investor's funding risk.
Neglecting double counting issues as a first approximation,  we may set
\beq \lab{deffcva}
\FRCVA^{1} = \FRA^{1}  +\CVA^{1} . 
\eeq
\ed

\brem
{\bf Additive adjustment}. Even in the valuation context, it is clear that funding costs are not really an additive adjustment, see \cite{PPB,NVA}. The industry applies an additive adjustment as a computationally feasible approximation. In our setup FRA is a risk measure, so it should not be simply added to the other adjustments coming from the risk neutral approach, such as CVA. Nonetheless, for the purpose of providing a single number expressing the deal profitability, we defined the FRCVA quantity above. This quantity has to be taken as indicative and not as a precise valuation quote. What retains real interest in our analysis, however, is the funding loss distribution and the related funding loss statistics. 
\erem

We will illustrate our definitions in the next section 
when examining the funding loss distribution of interest rate swaps. We now provide further details on the computation
of the funding loss. 

When computing the funding loss distribution, special attention must be paid to modelling the funding spread. 
The model must rightly reflect the way the financial institution finances
its activities, as well as the relationship between the Treasury department and the trading desk.
Also, the interplays between the funding requirement and the funding process must be fully understood. 

Let us start our discussion by distinguishing the cost of funding from the benefit of investing 
for the trading desk, as quoted by the treasury.

\bd
The {\it funding spread} $\phi^{1,+}_t$ for the investor is the instantaneous cost of borrowing one unit of cash at time $t$ from the treasury department for the trading desk of the investor.
\ed

\bd
The {\it investing spread} $\phi^{1,-}_t$ for the investor is the instantaneous benefit of lending one unit of cash at time $t$ to the treasury department for the trading desk of the investor.
\ed

We point out that if a trading desk is always net borrowing, giving back cash to the treasury will simply reduce lending, and the benefit will presumably be the same rate the treasury charges for borrowing.  This symmetry however could be broken by treasury policies taking into account maturity transformation, netting sets, incentives and other policies. 

We further introduce the Credit Support Annex (CSA) collateral funding rate. 

\bd
The {\it Credit Support Annex collateral funding rate} $\psi_t$ stands for the rate that is earned
by an obligor on its posted collateral as specified in the Credit Support Annex agreed upon with its
counterparty. 
\ed

For instance, the CSA funding rate on cash collateral is usually the overnight interest rate,
namely the effective Federal Funds rate for US dollars, TOIS for Swiss francs, EONIA for 
Euro and SONIA for British pounds. 

In the next definition we introduce the general concept of right-and-wrong way funding risk,
which reflects the dependence between the funding requirement and funding spreads. 

\bd \label{def2.15c}
The {\it investor's right} ({\it wrong}, resp.) {\it way funding risk} is the negative (positive, resp.) dependence between the investor's
funding spreads $\phi^{1}_t$ and the funding requirement process $f^{1}_t $.
\ed

In other words, there is wrong (right resp.) way risk for the investor when his funding requirements tend to increase
(decrease resp.) when its funding
spread $\phi^{1,+}$ deteriorates ( improves resp.), resulting in increasing (decreasing resp.) funding costs. 
But
there is also wrong (right resp.) way risk for the investor when his funding requirements, when negative,
tend to decrease 
(increase resp.) when its investing
spread $\phi^{1,-}$ deteriorates ( improves resp.), resulting in decreasing (increasing resp.) funding gains. 

In the same line of thought we introduce the right-or-wrong way systemic funding risk. 

\bd \label{def2.15c}
The {\it investor's right} ({\it wrong}, resp.) {\it way systemic funding risk} is the positive (negative, resp.) dependence between  
systemic illiquidity  and the funding requirement process $f^{1}_t$.
\ed

Put it differently, 
there is wrong-way systemic funding risk for the investor if its funding requirements increase
when funding resources are becoming 
scarce within the financial system due to heightened systemic illiquidity risk, resulting in an overall increase in funding costs.  
We do not illustrate specifically this concept in the article but will address it in further research. We will also need to provide a precise definition of systemic illiquidity risk. 

To conclude this section we list some examples of formalization of funding loss under the hypothesis of collateralization or in the 
absence of collateralization, with hedging assets traded in swapped form and 
assuming that the impact on the funding loss distribution of the early termination due to early default is negligible. 
We will assume these two conditions to hold throughout the paper. The simplifications these two assumptions provide allow us to stress the main conclusions
drawn by our analysis without going into too much detail. 

We should mention that these two assumptions, which can be found in part of the literature,
are debatable, and we will examine them further in future research. For the time being, we believe it is important to explain how hedging strategies funding costs are considered. 
When resorting to a Monte-Carlo simulation, it is possible to try to replicate the 
actual hedging strategy that would be conducted by the trading desk.
Since the sensitivities of the contracts,  
as well as the ones of the hedging instruments, 
 can be computed at each Monte-Carlo step, 
one can estimate the number of contracts that the trading desk would 
need to purchase in view of hedging. One can then deduce the 
actual cash necessary to purchase the hedging instruments.
Consecutively one can 
try to assess the funding costs not in light of 
its theoretical replication price but given this specific cash and hedging 
policy of the desk. If the desk envisages several scenarios of practical replication,
then the desk can decide to assess the funding loss distribution for each of these 
specific scenarios in order to get additional insight about the most appropriate strategy. 
This in turn shows that there is neither a unique 
funding loss distribution for the investor, nor a unique profitability profile for a given financial instrument.

\bp \lab{f1}
The funding loss of an uncollateralized contract, assuming that the hedging assets are traded in swapped form at no upfront payment, is given for the investor by
\be \label{fundfirst} \nonumber
\Phi^{1} = \int_0^{\tau^1 \wedge \tau^2 \wedge \bar{T} }  \, D(0,t) \, \bigg( \, \phi^{1,+}_t \, \big( -\bhP_t \big)^- - \phi^{1,-}_t \, \big( -\bhP_t \big)^+ \, \bigg) \, dt
\ee
\ep

The intuition behind this proposition is as follows. It is assumed that the funding costs come solely from the hedging process $H_t$. At any time t, the investor is replicating the counterparty contract market value including the potential loss upon default hence the investor holds $-\bhP_t$.
If $-\bhP_t$ is positive (resp. negative), the investor can invest (must borrow) the proceeds through the treasury department, having an instantaneous benefit
(resp. cost) of
$\phi^{1,-}_t \, \big( -\bhP_t \big)^+$ (resp. $\phi^{1,+}_t \, \big( -\bhP_t \big)^-$). Finally notice that, first, the investor does not need to borrow any additional cash, nor does he receive any proceedings associated with the purchase of the hedging instruments, since they are in swapped form,
and second, the hedging is perfect in the sense that there is no additional cost or benefit, on the event of counterparty default. This assumes somewhat unrealistically that default risk is perfectly hedged.  This implies that the replacement loss is null, as was hypothesized, which means in particular that there is no contagion effect at the first default / closeout.

We now consider the situation of a collateralized contract. 

\bp \lab{f2}

The funding loss of a collateralized contract, assuming that the hedging assets are traded in swapped form at no upfront payment,
is given for the investor by
\be \nonumber
\Phi^{1} = \int_0^{\tau^1 \wedge \tau^2 \wedge \bar{T} }  \, D(0,t) \, \bigg( \,  \psi_t  \, \big( -\bhP_t \big)^- - \psi_t  \, \big( -\bhP_t \big)^+ \, \bigg) \, dt
= \int_0^{\tau^1 \wedge \tau^2 \wedge \bar{T} }  \, D(0,t) \, \psi_t \, \bhP^{1}_t \, dt
\ee
\ep

The idea here is that the funding costs come both from the hedging process $H_t$ and the collateralization process $C_t$ .
At any time t, the investor is replicating the counterparty contract market value including the potential loss
upon default hence is holding $-\bhP_t$.
If $-\bhP_t$ is positive, the investor can use the proceeds as a collateral and 
hence receive the associated collateral funding rate. On the other hand, if 
$-\bhP_t$ is negative, the investor receives collateral whose 
amount is equal to the replication loss and hence does not need to borrow from
treasury. However the investor needs to finance the received collateral at the collateral funding rate.

It is important to stress that collaterals due to 
a trading desk are, in practice, not actually received by the desk but can be kept in the treasury
or back office. Also, the pricing of a stand alone contract in the presence
of collateralization typically neglects that collateralization occurs at the
level of the whole portfolio of contracts with the counterparty, which
implies that the knowledge of the contract marked-to-market 
value may not be enough so as to estimate the future exchange of  collaterals. 
These two comments imply that the actual collateral process within the financial institution
must be properly scrutinized. 

As a matter of illustration, we consider the 
funding costs of a contract where collateral is not 
passed to the desk. 

\bp \lab{f3}

The funding loss of a collateralized contract, assuming that the hedging assets are traded in swapped form at no upfront payment 
 and under the situation where collateral 
is not passed to the trading desk, is given for the investor by
\be \nonumber
\Phi^{1} = \int_0^{\tau^1 \wedge \tau^2 \wedge \bar{T} }  \, D(0,t) \, \bigg( \,  \phi^{1,+}_t \, \, \big( -\bhP_t \big)^- - \psi_t  \, \big( -\bhP_t \big)^+ \, \bigg) \, dt
\ee
\ep
In this case the funding costs come from both the hedging process $H_t$ and the collateralization process $C_t$.
At any time t, the investor is replicating the counterparty contract market value including the potential loss upon default, and   hence is holding $-\bhP_t$.
If $-\bhP_t$ is positive, the investor can use the proceeds as collateral, 
hence receiveing the associated collateral funding rate. Conversely, if 
$-\bhP_t$ is negative, the investor needs to borrow the full amount from the 
treasury. 

Finally, let us mention that it is sometimes argued that funding costs must be included into pricing since otherwise
the obligors would not agree on striking a deal as the addition of the (expected)
funding loss to the price (without funding costs), as seen from each obligor, could not match. 
It is true that funding loss introduces asymmetry. However, even in the replication analysis deal specific valuation measures and asymmetries may appear, see for example \cite{BPCCP}.

Moreover, we conjecture that if symmetry had to be present rigorously, then no deal would ever be striken in the market. 
Firstly, as it is shown for instance in \cite{DR}, counterparty credit risk can introduce price asymetry
anyway. Secondly, notice that it has been a common practice for banks to account for the funding loss 
as an indirect cost or benefit that impacts the profit $\&$ loss of trading desks hence
funding costs were to some extent already passed to the client. 
%To put it differently, the obligors
%do not have the same utility functions which allows for the respective contract apparent 
%profitabilities to be different.

\ssc{Weighted Cost of Funding Spread }

Funding and investing spreads are generally modelled by means of instantaneous spreads that are tied to the financial institution credit risk spread (\cite{BK}, \cite{C}, \cite{P}, \cite{PPB}). This practice does not always reflect the typical combination of short term and long term commitments an organisation may adopt  in order to finance its operations in the framework of a global 'Corporate Financing Strategy' set-up by the Finance or Treasury departments of the organisation itself. 
Also, possible 'macro-hedging' of funding costs by the financial institution is not always considered.  In fact, the precise modelling of the funding spread requires to dig into the intricacies of 
the organisation financing policy, and in particular the mandate of the treasury desk (see for instance Brigo et al. in \cite{MBP}). 
For instance, one would need to know whether the treasury desk operates as a 'profit center' that aims at maximising its profitability, if it applies different funding conditions depending on the riskiness of a contract and/or the credit riskiness of the counterparty etc.  The paper \cite{PPB} gives a couple of examples with different granularity for the funding strategy, but in general this problem is hardly addressed. 

The restricted level of generalisation we must keep here entitles us to propose a simplified {\it Weighted Cost of Funding Spread} ($\WCFS$) whose benefit is to be more general than the prevailing approaches in the literature whilst giving general intuitions about funding costs management. Before presenting the model, we need to state a few definitions.  

\bd
The {\it funding factor} $\Theta^1_t$ of the investor is the ratio at time $t$ of the amount of short-term funding of the investor with regard its total amount of funding. 
\ed

It is important to mention here that the approach ought to be natural to financial analysts as $\Theta$ can be considered as 
reflecting the proportion of working capital requirement per total amount of borrowing
of the bank. 
We now introduce the associated short-term and long-term funding spreads.

\bd
The {\it short-term funding spread}  $\phi^{s,1,+}_t$ for the investor is the instantaneous cost for the investor of borrowing cash short-term in the market at time~$t$.
\ed

In other words, the short-term treasury funding spread is equal to the average cost of funding restricted to the
short-term period. 

\bd
The {\it long-term funding spread} $\phi^{l,1,+}_t$ for the investor is the instantaneous cost for the investor of borrowing long-term cash in the market at time~$t$.
\ed

The short-term and long term funding spreads
typically correspond to some average of the repo and other short-term borrowings on one hand,
and bonds and other long-terms borrowings on the other hand.
We are now ready to introduce the Weighted Cost of Funding Spread.

\bd \lab{WFS}
The {\it Weighted Cost of Funding Spread} $\phi^{\Theta,1,+}_t$ for the investor  is the funding spread obtained by combining the short-term and long term funding spreads in proportion to the funding factor:
\beq 
\phi^{\Theta,1,+}_t = \Theta^1_t  \, \phi^{s,1,+}_t + \big(1 - \Theta^1_t \big) \, \phi^{l,1,+}_t.
\eeq
\ed

It is important to note that, as
a matter of simplification, we do not distinguish between the funding spreads of the trading desk and treasury (hence only refer to the "investor"). In other words, we assume that treasury passes to the trading desk the exact weighted 
funding cost corresponding to the investor's overall funding strategy. 

We proceed by introducing similar definitions for the funds invested by the financial institution. 

\bd
The {\it short-term investment spread}  $\phi^{s,1,-}_t$ for the investor is the instantaneous benefit for the investor of lending cash short-term in the market 
at time~$t$.
\ed

\bd
The {\it long-term investment spread} $\phi^{l,1,-}_t$ for the investor is the instantaneous benefit for the investor of lending cash long-term in the market  at time~$t$ .
\ed

\bd
The {\it investment factor} $\Gamma^1_t$ of the investor is the ratio at time $t$ of the amount of short-term investment of the investor with regard the total amount of investment. 
\ed

\bd
The {\it Weighted Cost of Investment Spread} $\phi^{\Gamma,1,-}_t$ for the investor  is the investment spread obtained by combining the short-term and long term investment spreads in proportion to the investment factor:
\beq 
\phi^{\Gamma,1,-}_t = \Gamma^1_t  \, \phi^{s,1,-}_t + \big(1 - \Gamma^1_t \big) \, \phi^{l,1,-}_t.
\eeq
\ed

It should be clear that the weighted spreads methodology facilitates the analysis of funding management
by giving boundary values for funding. For instance, it suffices to compute the
funding costs of a strictly positive funding exposure with the spread corresponding to the short-term funding spread (respectively the long-term
funding spread) to obtain a low bound (respectively high bound) forasmuch as long term funding is generally more costly than short term funding. 
We conjecture that the 'real' funding cost, or benefit, somehow lies 'in
between', i.e at a value that we can approximate by using the proper $\WCFS$/$\WCIS$ and  funding/investment factors $\Theta$/$\Gamma$.
Note in particular that the assessment of funding costs by resorting to the prevailing interest rate  
complemented by the CDS/bond basis for the financial institution
 would typically underestimate the funding loss.

%%%%%%%%%%%%%%%%%%%%%%%%%%%%%%%%%%%%%%%%%%%%%%%%%%%%%%%%%%%%%%%%%%%%%%%%%%%%%%%%%%%%%%
%%%%%%%%%%%%%%%%%%%%%%%%%%%%%%%%%%%%%%%%%%%%%%%%%%%%%%%%%%%%%%%%%%%%%%%%%%%%%%%%%%%%%%
\section{Interest Rate Swap Funding Loss and FRCVA} \label{sect2}
%%%%%%%%%%%%%%%%%%%%%%%%%%%%%%%%%%%%%%%%%%%%%%%%%%%%%%%%%%%%%%%%%%%%%%%%%%%%%%%%%%%%%%
%%%%%%%%%%%%%%%%%%%%%%%%%%%%%%%%%%%%%%%%%%%%%%%%%%%%%%%%%%%%%%%%%%%%%%%%%%%%%%%%%%%%%%

Typically, funding costs are computed within the industry by resorting to a spot funding curve
provided by treasury which allows to bootstrap forward funding spreads, the latter being used as proxies 
for the actual future funding spreads. 
In this section we rather compute and discuss the funding loss distribution as well as the FRCVA of uncollateralized interest rate swaps by resorting to stochastic weighted funding and investment spreads.

The interest rate swaps under scrutiny start in 1 year and are payer interest rate swaps with maturities 10 and 30 years, respectively.
The fixed legs compound annually with a rate of 3 $\%$ and the floating legs compound semi-annually. 
The notional is 1 million euros.
The investor and the counterparty, which we will also refer to as the "bank" and the "counterparty", are A and BBB rated 
(with "equivalent" flat spreads of 115 bp and 200 bp, respectively). 
Let us start by describing our modelling assumptions
before discussing numerical results. We should point out that the following examples have illustrative purpose and are not fully accurate in terms of market conventions and definitions. 

%%%%%%%%%%%%%%%%%%%%%%%%%%%%%%%%%%%%%%%%%%%%%%%%%%%%%%%%%%%%%%%%%%%%%
\ssc{Forward Interest Rate Swap}  
%%%%%%%%%%%%%%%%%%%%%%%%%%%%%%%%%%%%%%%%%%%%%%%%%%%%%%%%%%%%%%%%%%%%%

The holder at time t of a {\it payer forward interest rate swap} settled in arrears with tenor
$0 \le T_0, T_1, \ldots, T_m = T$ commits to swap at fixed dates $T_i, i= 1, 2, \ldots,m$ 
the floating Libor rate $ L(T_{i-1})$, resetting at $T_{i-1}$ for maturity $T_i$, against a fixed spread, denoted by $\kappa$, on a notional amount $N$.

The future, default-free, discounted cash flows at any time $t \in [0, T_1]$ are given by (neglecting accrual terms for simplicity)
\beq
\Pi(t,T) = N \, \sum_{i=2}^{m} D(t,T_i) \big(L(T_{i-1}) - \kappa \big)(T_i - T_{i-1}) \, .
\eeq
By the $i$th forward swap we mean the swap with the first reset date $T_{i-1}$ and the cash flows
at settlement dates $T_i, \ldots, T_m$.
The price at time 0 of the $i$th forward swap equals
\beq
S_0^{i,m} = N \, \EE \bigg[ \sum_{l=i+1}^{m} D(0,T_l) \big( L(T_{l-1}) - \kappa \big) \alpha_l \bigg]
= N \, B(0,T_{i}) - \sum_{l=i+1}^m c_l \, B(0,T_l),
\eeq
where we denote $c_l = N \,  \kappa \, \alpha_l $ for $l = i + 1, 2, \ldots, m-1$ and $c_m = N \,  ( 1 + \kappa ) \, \alpha_m$ with $\alpha_l = T_l - T_{l-1}$.
$B(0,T)$ denotes the zero coupon bond for maturity $T$ associated with the LIBOR curve. We point out that here we are assuming a single curve setup but we should generalize this to multiple curves.
Similarly the price at any time $t \le T_{i-1}$  of the forward swap is given by 
\beq
S_t^{i,m} = N \, \EE \bigg[ \sum_{l=i+1}^{m} D(t,T_l) \big( L(T_{l-1}) - \kappa \big) \alpha_l \bigg]
= N \, B(t,T_{i}) - N \, \sum_{l=i+1}^m c_l \, B(t,T_l).
\eeq

%%%%%%%%%%%%%%%%%%%%%%%%%%%%%%%%%%%%%%%%%%%%%%%%%%%%%%%%%%%%%%%%%%%%%
\ssc{Exposure Modelling Assumptions}  
%%%%%%%%%%%%%%%%%%%%%%%%%%%%%%%%%%%%%%%%%%%%%%%%%%%%%%%%%%%%%%%%%%%%%

The computation of the funding loss of an interest rate swap with stochastic funding spreads can be 
obtained through the joint Monte-Carlo simulation of the interest rate curve 
and the obligors' credit spreads. This approach not only
provides scenarios of the future marked-to-market 
values of the swap but gives also the associated funding spreads at any Monte-Carlo step, 
which allows for the swift computation of the funding loss distribution.

Although the computation is made under the historical measure $\P$, we use the same framework for computing 
approximated values of the credit valuation adjustment (that is usually done under $\Q$) for the interest rate swap in view of estimating the $\FRCVA$. 

We should mention that we approximate the
future marked-to-market values by the risk free values, that is we do not take the
future credit valuation adjustment or a possible expectation of funding losses due to the replacement close-out into account. The approximation avoids having to resort to more time consuming simulations such as Monte-Carlo of Monte-Carlos, sub-paths and the likes. 

Following on the lines of Jamshidian and Zhu \cite{FJYZ}, we model the interest rate term structure dynamics by calibrating $n$ zero coupon continuously compounded rates as lognormal exponential Orstein Uhlenbeck processes. This approach for the short rate would be similar to a Black-Karazinski model, but here we use it for the spot rate. Given the spot rates at a point in time we interpolate the whole zero coupon rates across maturities. 
More specifically, let us denote by $Z_i = R(t,T_i)$  the zero-coupon continuously-compounded spot interest rate at time $t$ for maturity $T_i$. We set
\begin{equation} \label{eq:Zifac}
Z_i(t) = z_i(t) \exp\left( X_i(t) - \frac{1}{2} \nu_i^2(t )\right) ,\,  i \in [1, \ldots, n] 
\end{equation}
where $z_i$ denotes the forward value of the zero coupon rate and $X_i$ is its
associated Ornstein-Uhlenbeck process, with variance $\nu_i^2$ given below:
\beq
dX_i(t) = - \lambda_i \, X_i(t)  \, dt + \sigma_i \, dW_i(t), \, i \in [1, \ldots, n]   \, ,
\eeq
where $\lambda_i $ and $\sigma_i $ stand for the mean reversion speed and the
volatility of the risk driver $X_i$, whilst $W_i$ denotes its associated Brownian motion.

We need to specify under which measure $W$ is a Brownian motion, which amounts to specify the expectation hypothesis we wish to adopt. If we work under the physical measure $\P$ as initially stated, then $W$ is a Brownian motion under $\P$. In such a case, we would have

\[ \EPP[Z_i(t) ] = z_i(t)  \]

so that $z_i$ would not be real forward rates but $\P$ expectations of future spot rates.  Using forward measures would bring us back to real forward rates. In general there will be a market price of risk affecting $z_i$ when compared with actual forward rates. In this first analysis we neglect the market price of risk estimation and we do not discuss expectation hypotheses in detail, similarly to \cite{FJYZ}. 

We recall that the Ornstein-Ulhenbeck process is normally distributed, so that we have 
\beq
X_i(t)  \equiv \mathcal{N}( 0, \nu_i^2(t) ) \, , X_i(0) = 0 
\eeq
with
\beq
\nu_i^2(t) = \sigma_i^2 \frac{1- \exp(-2  \, \lambda_i t)
}{  2 \, \lambda_i }.
\eeq

As is well known, it follows that the volatility is strictly increasing and bounded by the 
long term volatility $\sigma_{i,\infty}^2$ such that
\beq
\sigma_{i,\infty}^2 = \frac{ \sigma_i^2 }{ 2  \, \lambda_i } .
\eeq

Also, the risk driver $X_i(t)$, whose initial value is null, is given by 
\beq
X_i(t) = \exp(-\lambda_i \, t ) \int_0^t \exp(\lambda_i \, u ) \, \sigma_i \, dW_i(u) \, , X_i(0) = 0
\eeq
so that the zero coupon rate is given at any time $t$ by substituting this last formula in (\ref{eq:Zifac}). 

The correlation among the zero coupon rates' risk drivers is given by the following
 matrix $ ( \rho ) $ of quadratic covariations:
\beq
\rho_{i,j} \ dt = d \langle W_i \,, W_j \rangle_t  , \  \  \  (i,j) \in [1, \ldots, n]^2 \, .
\eeq

In addition, we assume that the credit spreads $\kappa^i$ of the obligors across maturities are modelled by shifting the initial
credit spread curve as a function of the evolution of the 5 years spread.
More specifically the $x$ years CDS spread is given by  : 
\beq
\kappa^{i,xY}_t = \kappa^{i,xY}_0   \frac{  \kappa^{i,5Y}_t }{ \kappa^{i,5Y}_0 }    .
\eeq
where $i = 1$ for the investor (resp. 2 for the counterparty). 

In turn we assume the dynamics of the 5 years credit spread to be linked  to the evolution of some credit indices by  beta projection.  More specifically, the 5 years spread is lognormally distributed :
\beq
\kappa^{i,5Y}_t = \kappa^{i,5Y}_0 \, \exp \big(   \sigma_i \, W_i(t) - \frac{1}{2} \, (\sigma_i)^2  t \big) , 
\eeq
and its risk driver $W_i$ is given as a function of the risk drivers $W^I_k$ of the credit indices and an idiosyncratic component $\epsilon_i$:
\beq
 W_i(t) = \sum_k \beta_k \, W^I_k(t) - \sqrt{ 1 -  \sum_k \beta^2_k}  \,\epsilon_i ,
\eeq
where the $\beta_k$ are obtained by historical regression against the evolution 
of the spreads credit indices. The latter are modelled as log-normal process.  

Now, in order to account for the fact that the obligors can fund their activities in different currencies, 
we assume that the investor and the counterparty fund their operations in EUR and USD, respectively. 
The interest rate model we previously described is used for simulating both the 
EUR and the USD interest rate term structures. The exchange rate $\eta$ between EUR and USD is
modeled as a log-normal processes : 
\beq
\eta^{EUR,USD}_t = \eta^{EUR,USD}_0 \exp \big(   \sigma W^{\eta}(t) - \frac{1}{2} \,   (\sigma)^2 t \big) . 
\eeq

Clearly, one benefit of this approach is the ease with which one can construct the correlation structure between the diffusion
risk drivers and hence the risk factors. Formally, one can extend the correlation matrix to:
\beq
\rho_{i,j} \  dt =  d \langle W_i \,, dW_j  \rangle_t, \ \ \    (i,j) \in [1, \ldots, n +  k + 1]^2 \, .
\eeq
where we conveniently re-indexed the shocks $W^I$ and $W^\eta$, and where $n$ and $k$ denote the number of simulated zero coupon rates and 
the number of drivers used in the diffusion of each of the obligors' 5 year spreads, in addition
to the FX risk driver.

We should point out that in this first sketchy analysis we are including different currencies without analyzing rigorously 
the change of measure effects on default intensities and the FX effects on collateral, for which we refer to \cite{FST}.

%%%%%%%%%%%%%%%%%%%%%%%%%%%%%%%%%%%%%%%%%%%%%%%%%%%%%%%%%%%%%%%%%%%%%
\ssc{Funding Modelling Assumptions}  
%%%%%%%%%%%%%%%%%%%%%%%%%%%%%%%%%%%%%%%%%%%%%%%%%%%%%%%%%%%%%%%%%%%%%

At each Monte Carlo step, the price of each interest rate swap can be obtained analytically 
by means of the zero rates interest rate curve, that give us immediately all the relevant discount bonds needed to value the swap. 
%\beq
%S_t^{i,m} 
%= N \, B(t,T_{i-1}) - N \, \sum_{l=i+1}^m c_l \, B(t,T_l) = N \, \exp\big( - Z(t,T_{i-1}) \big) - N \, \sum_{l=i+1}^m c_l \, \exp\big( - Z(t,T_l) \big) \, .
%\eeq

In turn, the 
funding loss for the investor can be computed, again assuming that the hedging assets are traded in swapped form at no upfront payment, by
applying the equation
\be \nonumber
\Phi^{1} = \int_0^{\tau^1 \wedge \tau^2 \wedge \bar{T} }  \, D(0,t) \, \bigg( \, \phi^{\Theta,1,+}_t \, \big( -\bhP_t \big)^- - \phi^{\Gamma,1,-}_t \, \big( -\bhP_t \big)^+ \, \bigg) \, dt.
\ee
Assuming for simplicity that the counterparty is default free,
the funding loss distribution reads
\be \nonumber
\Phi^{1} = \int_0^{\tau^1  \wedge \bar{T} } \, D(0,t)  \, \bigg( \,  \phi^{\Theta,1,+}_t \, \big( -\bhP_t \big)^- - \phi^{\Gamma,1,-}_t \, \big( -\bhP_t \big)^+ \bigg) \, dt.
\ee
The previous equation puts into sharp focus the dependency of the funding loss on the investor's own probability of default. 
In the situation where the investor benefits also from a high credit quality, the formula can be approximated 
further for relatively short term financial instruments by
\be \nonumber
\Phi^{1} \simeq \int_0^{\bar{T} } \, D(0,t) \, \bigg( \,  \phi^{\Theta,1,+}_t \, \big( -\bhP_t \big)^- - \phi^{\Gamma,1,-}_t \, \big( -\bhP_t \big)^+ \bigg) \, dt.
\ee
Notice that in practice the termination date of the financial instrument does not necessarily correspond to its actual maturity.
 Banks tend to account for the expected termination date so as not to charge a funding adjutment that could prove to be uncompetitive.

In order to estimate the funding spread process so as to obtain the full distribution of the funding loss,
we resort to the weighted spread methodology we introduced in the previous chapter. 
As a matter of simplification,  we assume that the Weighted Cost of Funding Spread and the Weighted Cost of Investment Spread
are equal and in particular $\Gamma  = \Theta$.  Now, we specify the $\WCFS$, 

\beq 
\phi^{\Theta,i,+}_t = \Theta^i_t  \, \phi^{s,i,+}_t + \big(1 - \Theta^i_t \big) \, \phi^{l,i,+}_t \, ,
\eeq

with

\beq 
\phi^{s,i,+}_t  = Z^{3Y}_t + \kappa^{i,3Y}_t \, , \phi^{l,i,+}_t =  Z_t^{10Y} + \kappa^{i,10Y}_t, 
\eeq

where $Z^T_t$ denotes the zero coupon rate of maturity T years and $ \kappa_i $ stands for the obligor spread at time t, with $i=1$ ( resp. $i=2$ ) for the investor (respectively the counterparty). In other words, we assume that the 3 years (10 years respectively) spread, complemented by the 3 years CDS spread (10 years CDS spread respectively), gives a good representation of the short term (respectively long term) borrowing cost of the obligors. 
Notice that the funding rate depends on the funding currency of the obligor. For instance, for the investor, the equality reads

\beq 
\phi^{s,1,+}_t  = Z^{EUR,3Y}_t + \kappa^{1,3Y}_t \, , \phi^{l,1,+}_t = Z_t^{EUR,10Y} + \kappa^{1,10Y}_t \, .
\eeq
%%%%%%%%%%%%%%%%%%%%%%%%%%%%%%%%%%%%%%%%%%%%%%%%%%%%%%%%%%%%%%%%%%%%%
\ssc{Numerical Implementations and Representations}  
%%%%%%%%%%%%%%%%%%%%%%%%%%%%%%%%%%%%%%%%%%%%%%%%%%%%%%%%%%%%%%%%%%%%%

% Marked-to-Market value of approximately 120,000 euros and 200,000 euros for 10 and 30 years swaps
% \includegraphics[width=1.0\linewidth]{30YSwapEPEENE_.png}

\begin{figure}[!h]
\centering
\begin{minipage}{.5\textwidth}
  \centering
  \includegraphics[width=1.0\linewidth]{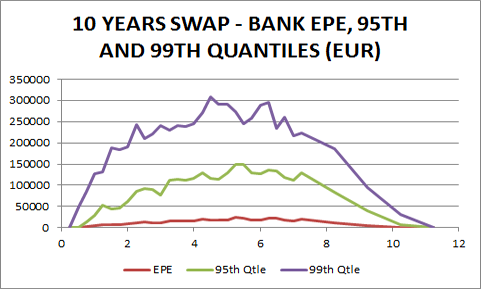}
  \captionof{figure}{ Bank - 10 Years Swap}
  \label{fig:test2}
\end{minipage}%
\begin{minipage}{.5\textwidth}
\centering
  \includegraphics[width=1.0\linewidth]{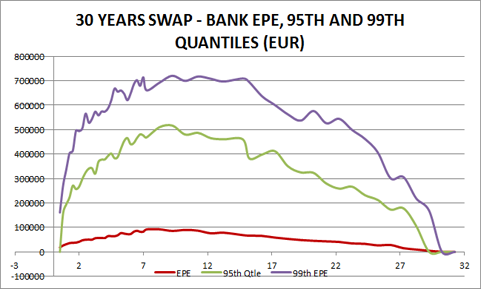}
  \captionof{figure}{\ \  Bank - 30 Years Swap}
  \label{fig:test2.2}
\end{minipage}
\end{figure}

We start our discussion by representing risk metrics for the exposures of the swaps. 
Figure \ref{fig:test2}  and  Figure \ref{fig:test2.2} 
display the Expected Positive Exposure, 95th percentile and 
99th percentile exposures of the 10 years and 30 years swaps, from the Bank point of view.

\begin{figure}[!h]
\centering
\begin{minipage}{.5\textwidth}
  \centering
  \includegraphics[width=1.0\linewidth]{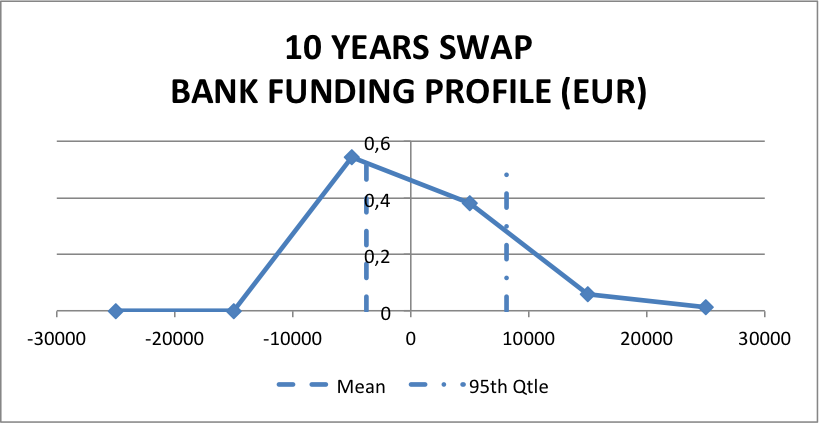}
  \captionof{figure}{Bank - 10 Years Swap\\ Funding - PDF:  $\Theta = 1 $}
  \label{fig:test3}
\end{minipage}%
\begin{minipage}{.5\textwidth}
  \centering
  \includegraphics[width=1.0\linewidth]{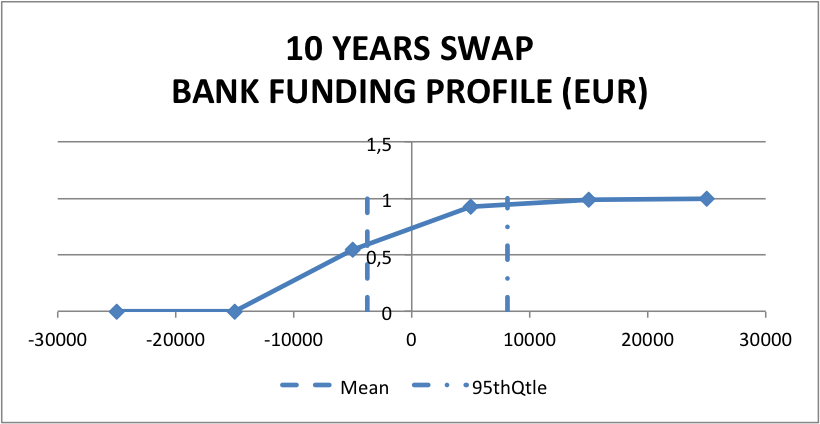}
  \captionof{figure}{Bank - 10 Years Swap\\ Funding - CDF:  $\Theta = 1 $}
  \label{fig:test4}
\end{minipage}
\end{figure}

Next, Figures \ref{fig:test3} and \ref{fig:test4} exhibit the funding distributions of the investor for the 10 years swap, 
under the hypothesis that the funding costs are function of the short term funding only ($\Theta=1$).
The probability densities as well as the cumulated distribution functions are given. We also show the mean and the $95\%$th percentile.
The mean is negative since the exposure is mostly negative for the bank. 
More precisely, in this example the bank expects to reap a general funding benefit of 3700 euros. With different exposure profiles and different funding rates, we could easily get an expected cost instead of an expected benefit. Our numerical studies later will still concern this example, so in Tables \ref{data13} and \ref{data16} we will see mostly a benefit situation on average, but this is just an example. However, we should not look just at the expectation. The funding loss distribution support is both negative and positive, resulting in a benefit or a cost for the investor in different scenarios,  and exhibits a relatively large variance. 
Indeed, in approximately 20$\%$ of scenarios, total contract funding happens to be a cost, as it can be seen 
thanks to the cumulated loss distribution. In particular, the 95th percentile represents a potential loss of 
8,000 Euros. 

\begin{figure}[!h]
\centering
\begin{minipage}{.5\textwidth}
  \centering
  \includegraphics[width=1.0\linewidth]{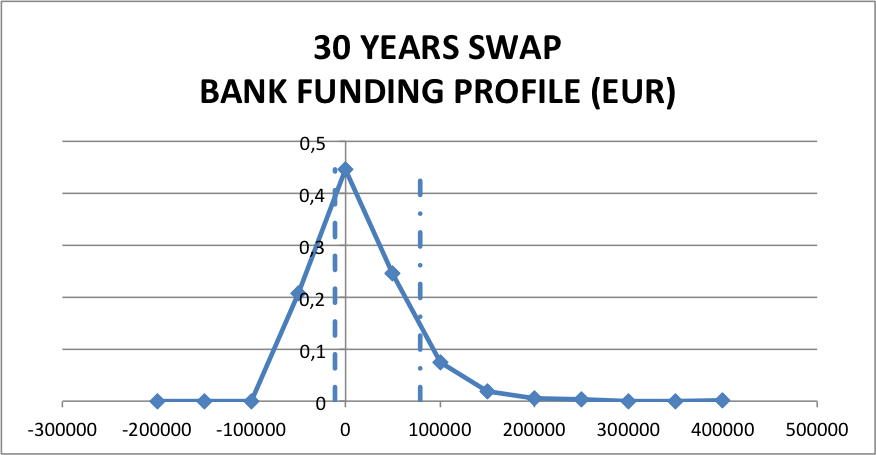}
  \captionof{figure}{Bank - 30 Years Swap\\ Funding  PDF:  $\Theta = 1 $}
  \label{fig:test5}
\end{minipage}%
\begin{minipage}{.5\textwidth}
  \centering
  \includegraphics[width=1.0\linewidth]{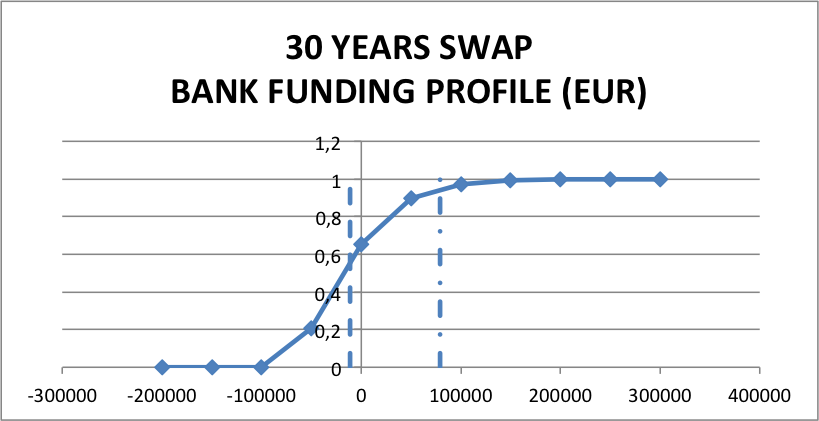}
  \captionof{figure}{Bank - 30 Years Swap\\ Funding CDF: $\Theta = 1 $}
  \label{fig:test6}
\end{minipage}
\end{figure}

Figures \ref{fig:test5} and \ref{fig:test6} provide similar statistics regarding the funding loss distribution
of the 30 years swap. The  mean indicates a benefit of 11,000 euros, whilst the variance 
of the distribution as well as the probability of an actual loss have increased compared to the 
10 years case. The bank can suffer a loss with a probability of approximately 30$\%$, 
and the 95th percentile reaches 80,000 Euros, that is roughly 8 times the mean. 

\begin{figure}[!h]
\centering
\begin{minipage}{.5\textwidth}
  \centering
  \includegraphics[width=1.0\linewidth]{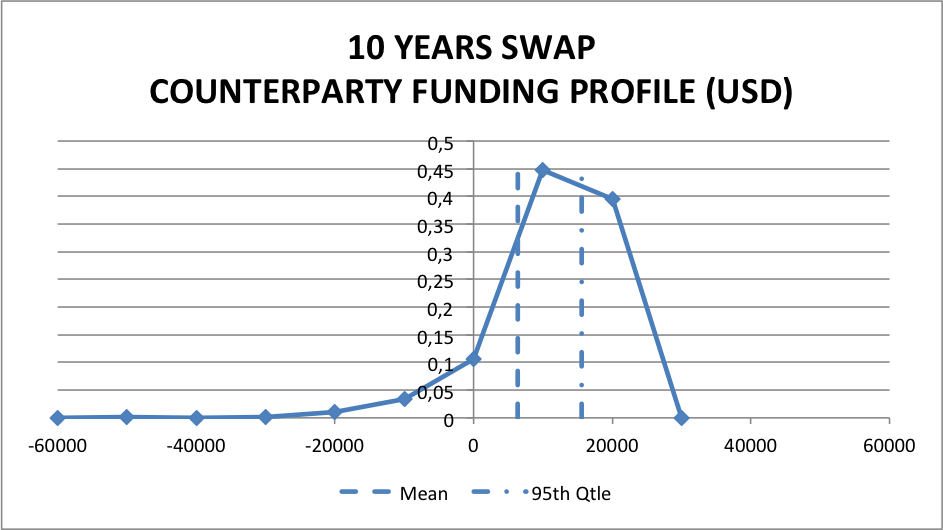}
  \captionof{figure}{Counterparty - 10 Years Swap\\ Funding PDF: $\Theta = 1 $}
  \label{fig:test7}
\end{minipage}%
\begin{minipage}{.5\textwidth}
  \centering
  \includegraphics[width=1.0\linewidth]{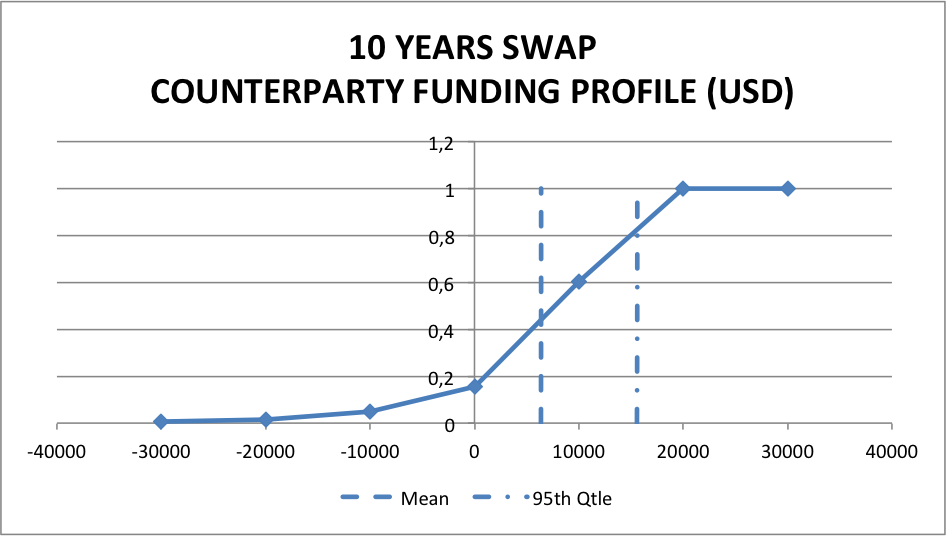}
  \captionof{figure}{Counterparty - 10 Years Swap\\ Funding CDF:  $\Theta = 1 $}
  \label{fig:test8}
\end{minipage}
\end{figure}

\begin{figure}[!h]
\centering
\begin{minipage}{.5\textwidth}
  \centering
  \includegraphics[width=1.0\linewidth]{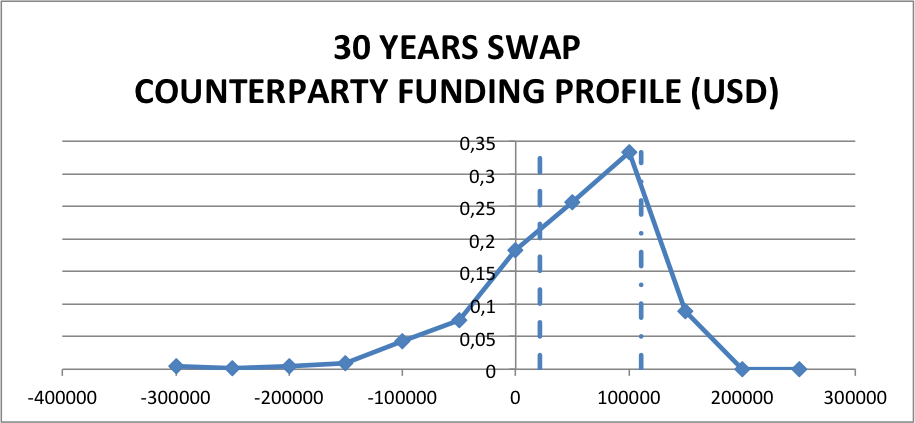}
  \captionof{figure}{Counterparty - 30 Years Swap\\ Funding PDF: $\Theta = 1 $ }
  \label{fig:test9}
\end{minipage}%
\begin{minipage}{.5\textwidth}
  \centering
  \includegraphics[width=1.0\linewidth]{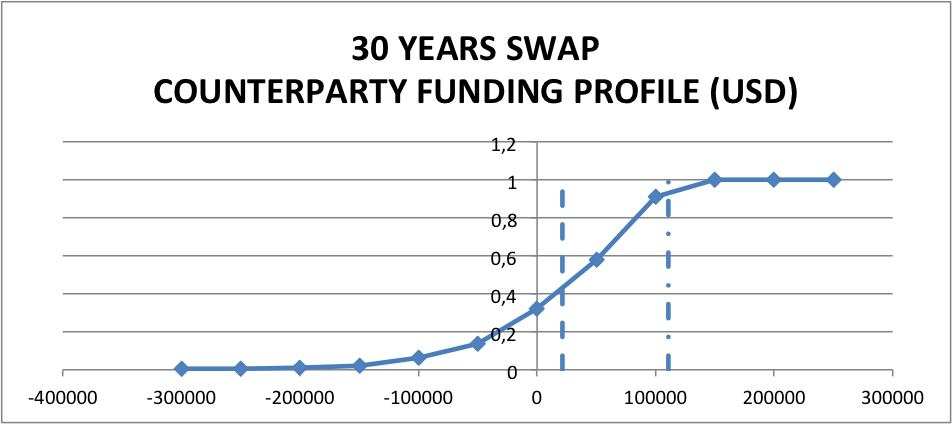}
  \captionof{figure}{Counterparty - 30 Years Swap\\ Funding CDF:  $\Theta = 1 $}
  \label{fig:test10}
\end{minipage}
\end{figure}

Figures  \ref{fig:test7}-\ref{fig:test10}  provide the funding statistics for the 10 years and 30 years swaps from the counterparty
point of view. As expected, the means are positive since the exposures are generally positive. More specifically, 
for the 10 years swap, the  mean equals 6400 USD and the 95th percentile is 16,000 USD, whilst for the 30 years swap,  
 $\FRA$ = 21,000 USD and 95th percentile is  110,000 USD. Again the risk in absolute and relative terms tends to increase
with swap duration. 

We now represent the funding distributions under the hypothesis that long term funding contributes 
to total funding in the proportion of 40 $\%$ ($\Theta = 0.6$). 
Long term funding being typically more expensive than short term funding, 
it is expected that when the bank reaps a benefit for a given Monte-Carlo scenario, the latter will be greater than under the 'short-term funding only' case,
whereas, when the banks suffers a cost, the latter will be larger as well. In turn, one expects the variance of the funding 
loss distribution to increase, all the more so since the addition of the long term funding risk introduces another parameter 
in our model. Also, the behaviour of the mean is not as clear cut as one might expect, since it depends on the relative impact of the 'increased positive scenarios' 
and 'increased negative scenarios' (in absolute terms). 
Indeed one can observe first an increase in loss distributions' variances, and second the absence of a clear  mean trend. 
In particular, the 95th percentile of the loss distributions 
goes from 16,000 to 22,000 euros, and from 110,000 to 225,000 euros, for the 10 years swap and the 30 years swap, respectively. 
The mean benefit of the 10 years swap increases significantly from 3,700 to 10,000 euros, whilst the mean of the 30 years swap stays stable at 14,000 euros against 11,000 euros. However, the take-away point is that risk increases when one considers the impact of long-term funding.
\begin{figure}[!h]
\centering
\begin{minipage}{.5\textwidth}
  \centering
  \includegraphics[width=1.0\linewidth]{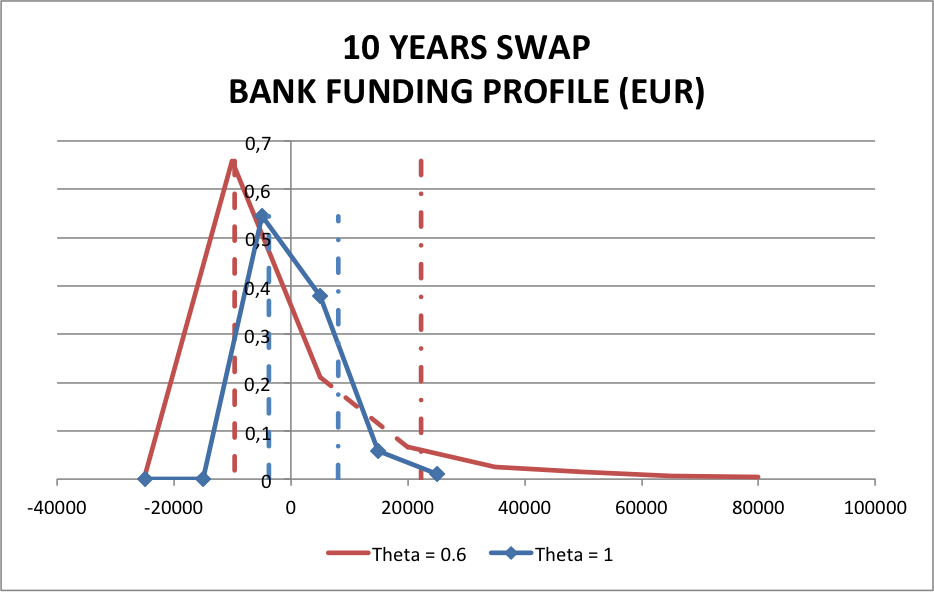}
  \captionof{figure}{Bank - 10 Years Swap\\ Funding PDF: $\Theta \in \{1 , 0.6 \}$}
  \label{fig:test11}
\end{minipage}%
\begin{minipage}{.5\textwidth}
  \centering
  \includegraphics[width=1.0\linewidth]{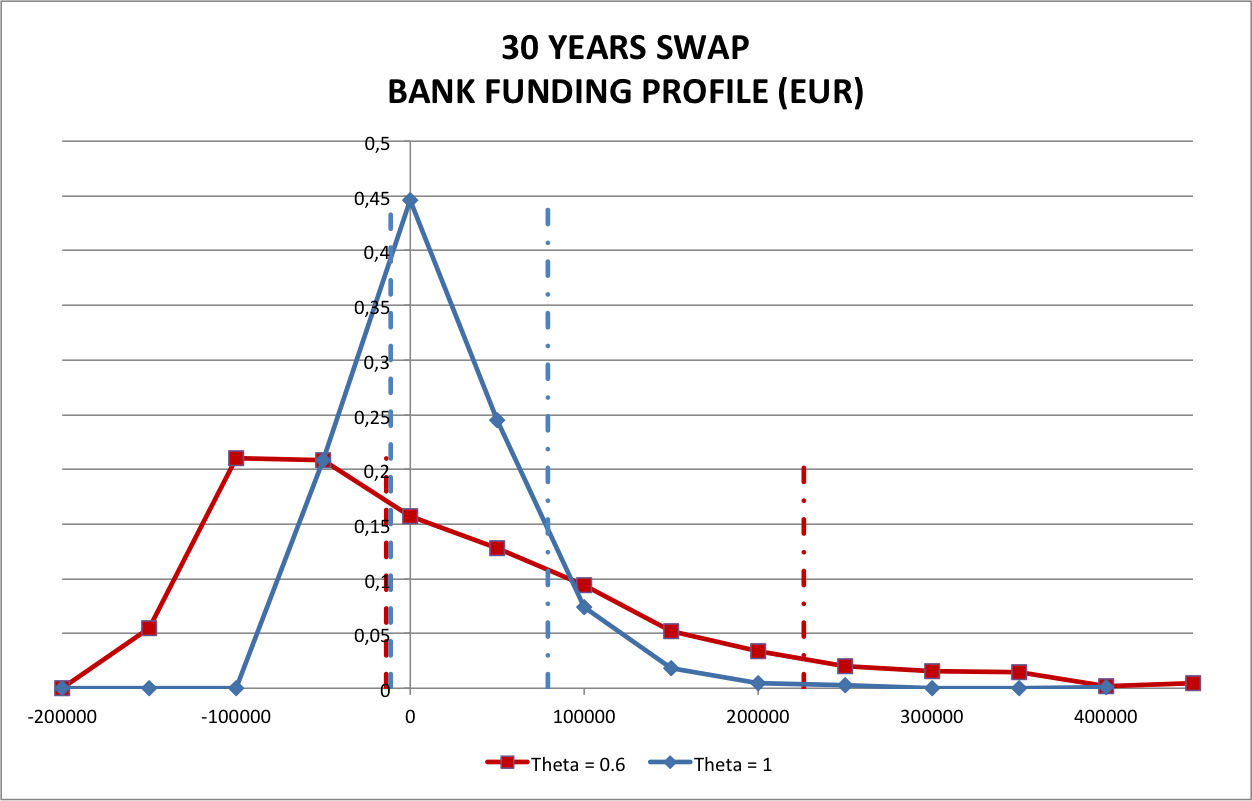}
  \captionof{figure}{Bank - 30 Years Swap\\ Funding PDF: $\Theta \in \{1 , 0.6 \}$ }
  \label{fig:test12}
\end{minipage}
\end{figure}

\begin{figure}[!h]
\centering
\begin{minipage}{.5\textwidth}
  \centering
  \includegraphics[width=1.0\linewidth]{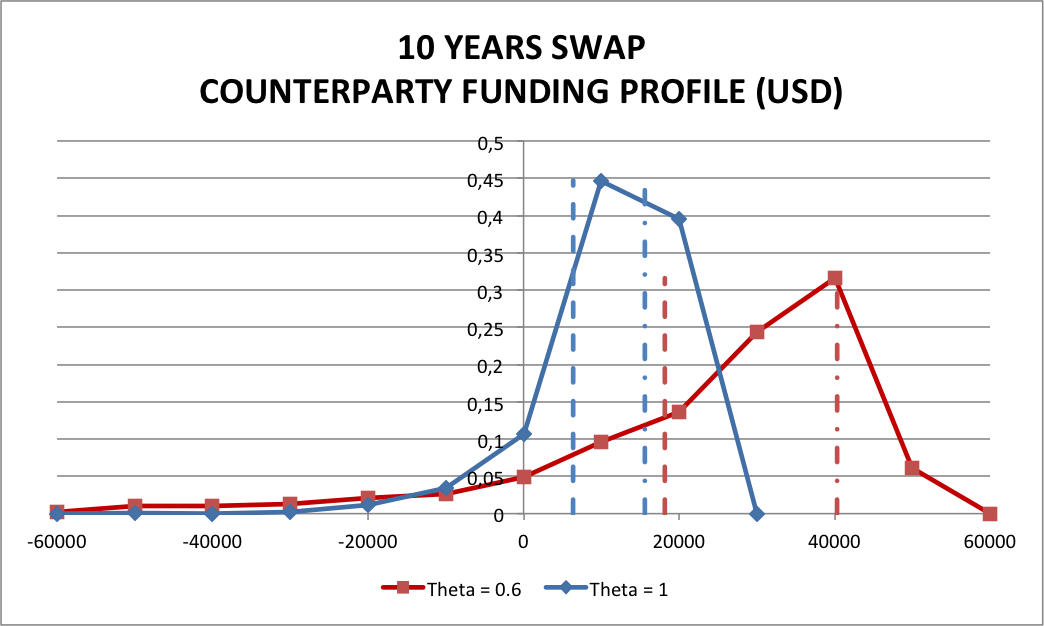}
  \captionof{figure}{Counterparty - 10 Years Swap\\ Funding PDF: $\Theta \in \{1 , 0.6 \}$}
  \label{fig:test13}
\end{minipage}%
\begin{minipage}{.5\textwidth}
  \centering
  \includegraphics[width=1.0\linewidth]{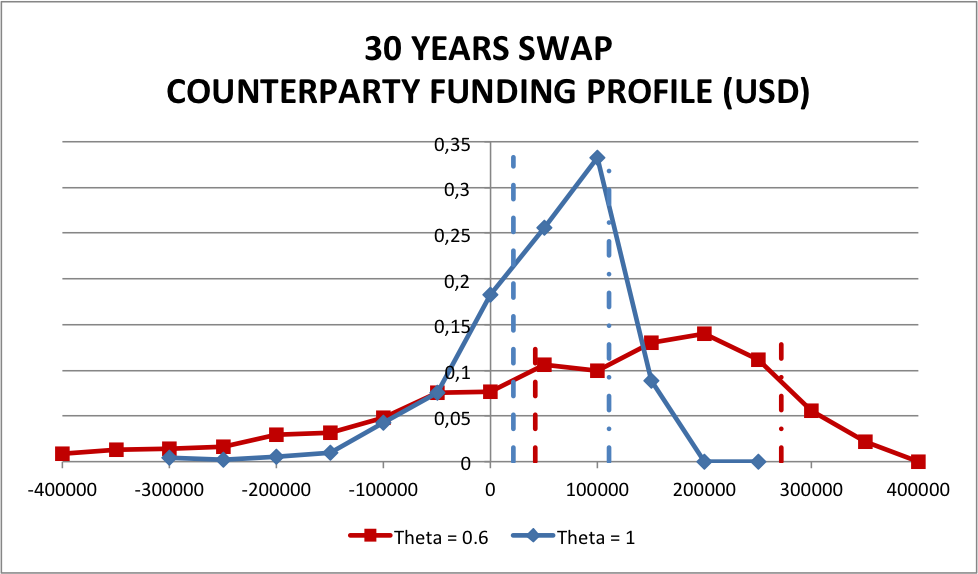}
  \captionof{figure}{Counterparty - 30 Years Swap\\ Funding PDF: $\Theta \in \{1 , 0.6 \}$ }
  \label{fig:test14}
\end{minipage}
\end{figure}

Figures \ref{fig:test13} and \ref{fig:test14} exhibit similar features. 
The 95th percentiles  of the 10 years and 30 years swaps are multiplied by 2.5, from 16,000 to 40,000 USD, and from 110,000 to 270,000 USD, for the 10 years swap and the 30 years swap, respectively. 

\begin{figure}[!h]
\centering
\begin{minipage}{.5\textwidth}
  \centering
  \includegraphics[width=1.0\linewidth]{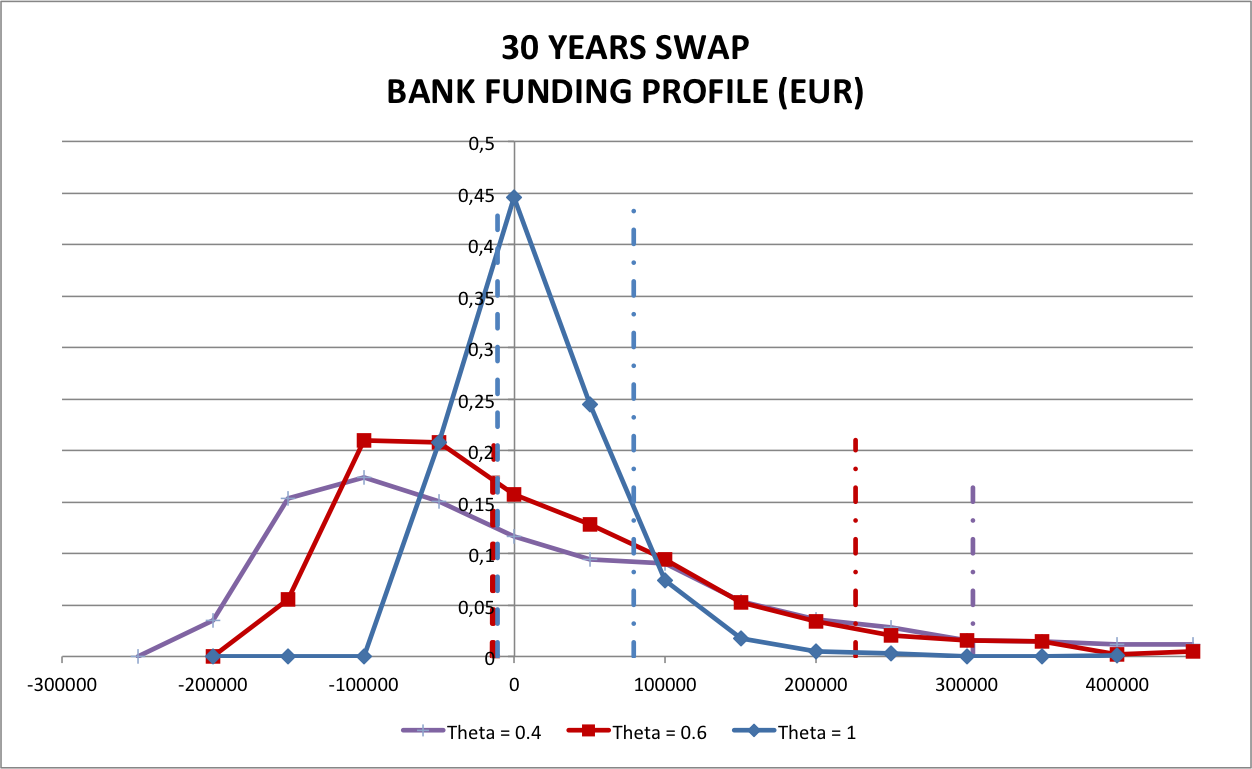}
  \captionof{figure}{Bank - 30 Years Swap\\ Funding PDF: $\Theta \in \{1 , 0.6, 0.4 \}$}
  \label{fig:test15}
\end{minipage}%
\begin{minipage}{.5\textwidth}
  \centering
  \includegraphics[width=1.0\linewidth]{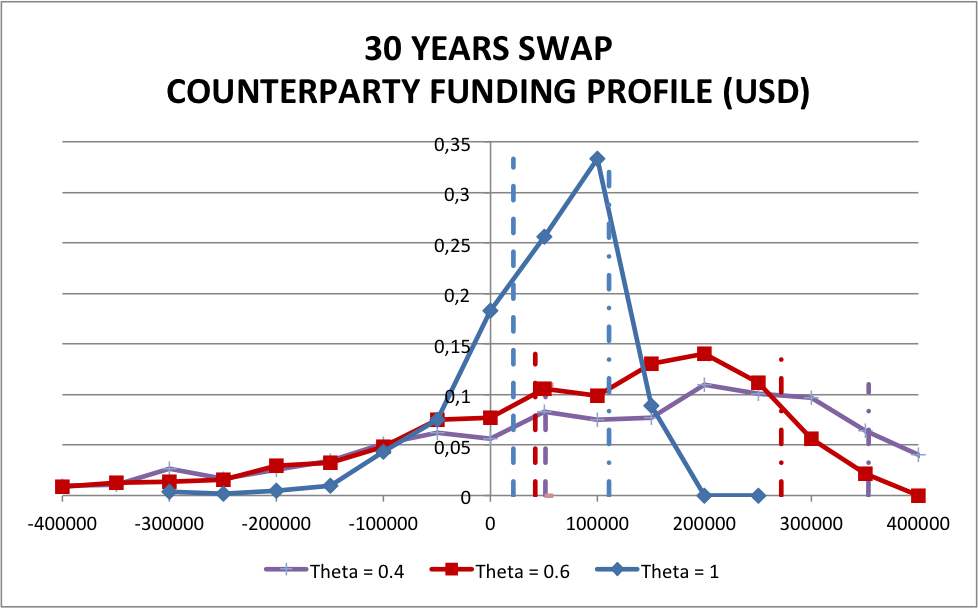}
  \captionof{figure}{Counterparty - 30 Years Swap\\ Funding PDF: $\Theta \in \{1 , 0.6, 0.4 \}$ }
  \label{fig:test16}
\end{minipage}
\end{figure}

These comments are confirmed when complementing our analysis with the funding loss distributions 
for the bank and the counterparty when the 
short term funding is limited to $40\%$ of total funding ($\Theta=0.4$). 
As a matter of illustration, Figures 
$\ref{fig:test15}$ and $\ref{fig:test16}$ show our results for the 30 years swap. 
Again, they emphasize that, although the  mean is relatively stable in absolute terms, 
this should not give a false sense of security as the 95th percentile is very sensitive 
to the weight of long term funding.

\begin{figure}[!h]
\centering
\begin{minipage}{.5\textwidth}
  \centering
  \includegraphics[width=1.0\linewidth]{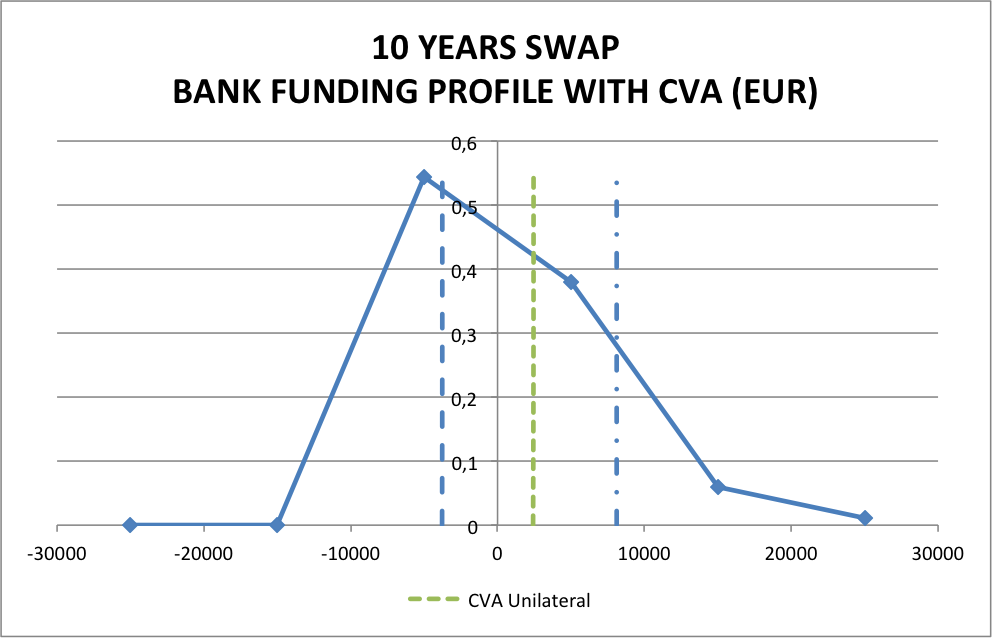}
  \captionof{figure}{Bank - 10 Years Swap\\ Funding Profile with CVA: $\Theta = 1$}
  \label{fig:test18}
\end{minipage}%
\begin{minipage}{.5\textwidth}
  \centering
  \includegraphics[width=1.0\linewidth]{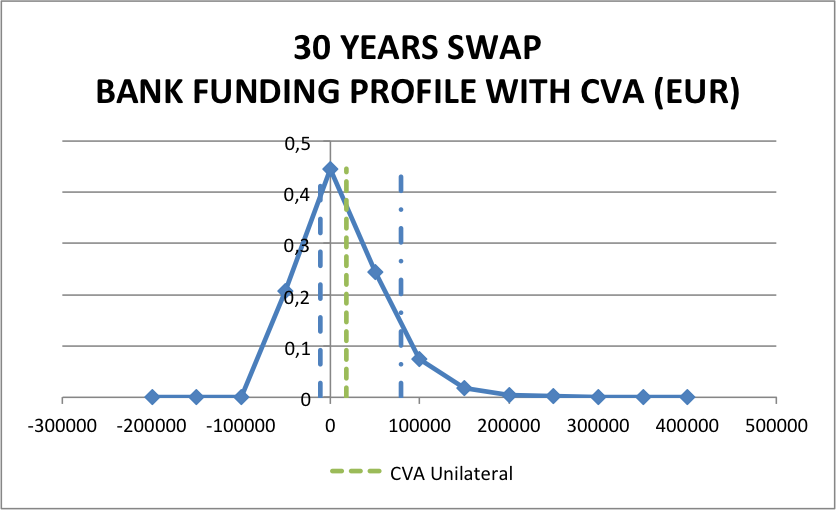}
  \captionof{figure}{ Bank - 30 Years Swap\\ Funding Profile with CVA: $\Theta = 1$ }
  \label{fig:test19}
\end{minipage}
\end{figure}

\begin{figure}[!h]
\centering
\begin{minipage}{.5\textwidth}
  \centering
  \includegraphics[width=1.0\linewidth]{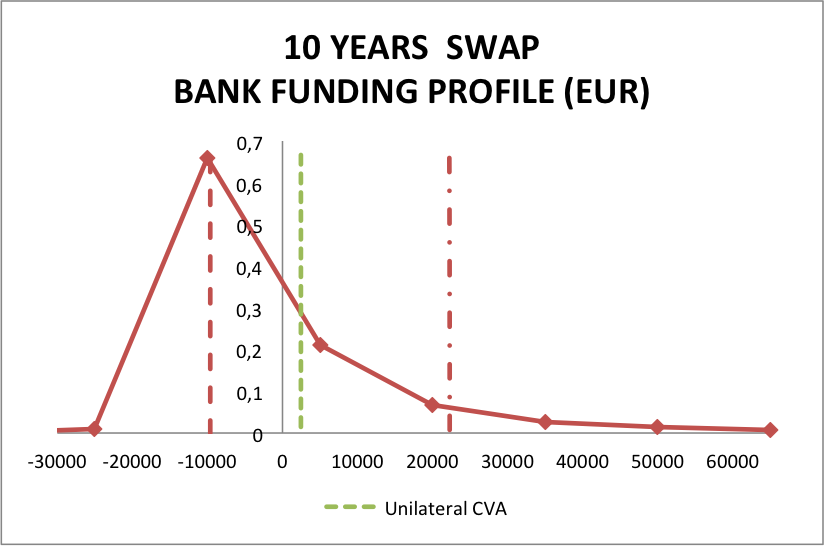}
  \captionof{figure}{Bank - 10 Years Swap\\ Funding Profile with CVA: $\Theta = 0.6$}
  \label{fig:test20}
\end{minipage}%
\begin{minipage}{.5\textwidth}
  \centering
  \includegraphics[width=1.0\linewidth]{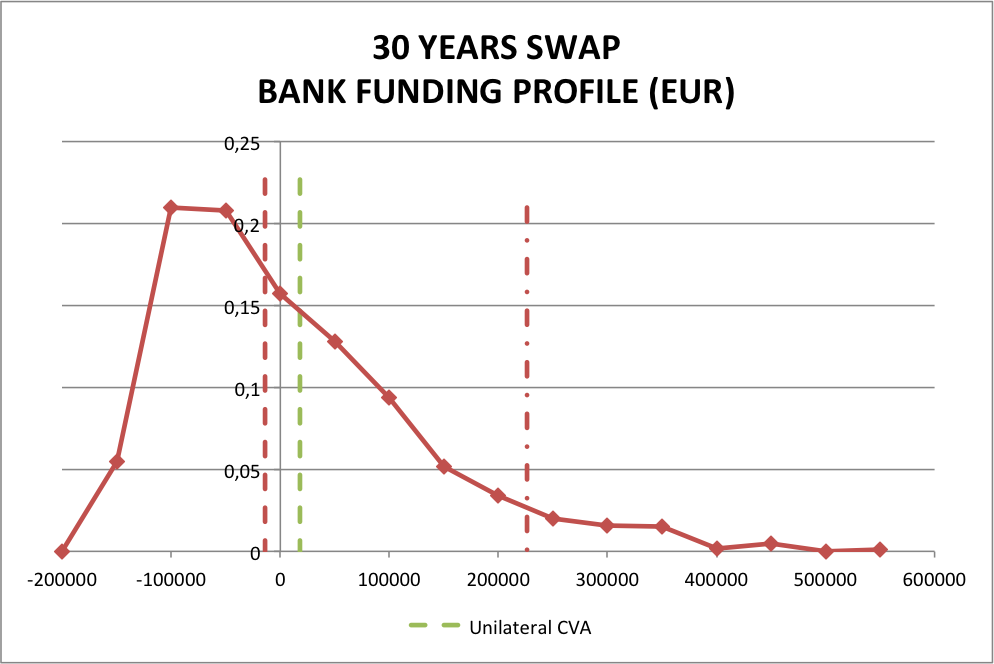}
  \captionof{figure}{ Bank - 30 Years Swap\\ Funding Profile with CVA: $\Theta = 0.6$}
  \label{fig:test21}
\end{minipage}
\end{figure}

Let us now assume that the Funding Benefit Adjustment and the Debt Valuation Adjustment are equal 
 and add the (unilateral) Credit Valuation Adjustment to our examination.
%More precisely, we consider both 'unilateral $\CVA^i$', which correspond to the case where obligor $i$ cannot default, and 'bilateral $\CVA^i$', where both %obligors can potentially default before contract termination. 
%The probabilities of survival of both the bank and its counterparty are given in Figure $\ref{fig:test17}$.
%\begin{figure}[!h]
%\centering
%\begin{minipage}{.5\textwidth}
 % \centering
%  \includegraphics[width=1.0\linewidth]{BankCtpySurvivalProbabilities.png}
 % \captionof{figure}{Bank and Counterparty Survival Probabilities as a function of Years}
  %\label{fig:test17}
%\end{minipage}%
%\end{figure}
Figures $\ref{fig:test18}$ and $\ref{fig:test19}$ display %both unilateral and bilateral
the credit valuation adjustment as well as the funding loss distribution corresponding to the 'short-term funding
only case', as seen from the bank. Let us mention here that, for ease of explanation, we do not discuss 
the counterparty case, as our observations lead to similar comments. 
The unilateral $\CVA^1$ for the 10 years (resp. 30 years) swap amounts to 2,500 euros 
(resp. 18,000 euros).
% and the bilateral $\CVA^1$ for the 10 years (resp. 30 years) swap equals -3,000 euros (resp. -9,000 euros). 
%As expected, the orders of magnitude of $\CVA$ and $\FRA$ are similar, and all the more so for bilateral $\CVA$ and $\FRA$ within our framework. 
We also show in Figures $\ref{fig:test20}$ and $\ref{fig:test21}$ these statistics in the $\Theta = 0.6$ case. 
Notice that, in the 10 years swap case, the mean increases significantly.

We conclude this chapter by discussing funding risk credit valuation adjustment.
Consider the following
{\bf $\mathcal{M}$ Hypothesis}: the mean of the funding
loss distribution is a suitable risk statistics to express  funding risk.

% Tables  $\ref{data12}$  and  $\ref{data13}$
 Table  $\ref{data13}$ gives
the computation of the $\FRCVA$, under the $\mathcal{M}$ Hypothesis,
for the swaps without long term funding ($\Theta = 1$) as seen by the investor. % , with bilateral or unilateral $\CVA$.
We also consolidate our results for $\Theta = 0.6$ in 
% Tables $\ref{data15}$  and  
Table $\ref{data16}$.
Let us stress once again that the $\mathcal{M}$ Hypothesis  is not suitable in general.
The associated $\FRA$ could nevertheless be interpreted as some "low bound adjustment"
since it does not account for the possible cost of mitigation strategies
of the funding cost, such as for instance the possible allocation of dedicated 
capital.

%\begin{table}[!h]
%\begin{center}
%\begin{tabular}{|c|c|c|}
%\hline
%$\mbox{Bank}$&$ \mbox{10 Years Swap}$ &$\mbox{30 Years Swap}$\\
%\hline  
%\mbox{MtM} & - 120,000 & - 200,000 \\
%\hline  
%\mbox{ Bilateral $\CVA$}	& -3,000&-9,000\\
%\mbox{$\FRA$ - $\mathcal{M}$ Hypothesis }	&-3,700& -11,000 \\
%\hline
%\mbox{$\FRCVA$ } & -6,700&  -20,000 \\
%\hline
%\mbox{MtM with $\FRCVA$ - Bilateral CVA} & -126,700 & -220,000 \\
%\hline
%\end{tabular}
%\end{center}\caption{ $\FRCVA$ of the Bank - Bilateral $\CVA$  - $\theta = 1$ - $\mathcal{M}$ Hypothesis }
%\label{data12}
%\end{table}

\begin{table}[!h]
\begin{center}
\begin{tabular}{|c|c|c|}
\hline
$\mbox{Bank}$&$ \mbox{10 Years Swap}$ &$\mbox{30 Years Swap}$\\
%\hline  
%\mbox{MtM} & - 120,000 & - 200,000 \\
\hline  
\mbox{ $\CVA$}	& 2,500	&18,000\\
\mbox{$\FRA$ - $\mathcal{M}$ Hypothesis  }	&-3,700& -11,000 \\
\hline
\mbox{$\FRCVA$ } & -1,200&  7,000 \\
\hline
%\mbox{MtM with $\FRCVA$ - Bilateral CVA} & -121,200 &  -193,000 \\
%\hline
\end{tabular}
\end{center}\caption{ $\FRCVA$ of the Bank - $\Theta = 1$ - $\mathcal{M}$ Hypothesis - Results quoted in Euros}
\label{data13}
\end{table}

%\begin{table}[!h]
%\begin{center}
%\begin{tabular}{|c|c|c|}
%\hline
%$\mbox{Bank}$&$ \mbox{10 Years Swap}$ &$\mbox{30 Years Swap}$\\
%\hline  
%\mbox{MtM} & - 120,000 & - 200,000 \\
%\hline  
%\mbox{ Bilateral $\CVA$}	& -3,000&-9,000\\
%\mbox{$\FRA$  - $\mathcal{M}$ Hypothesis }	&-10,000& -14,000 \\
%\hline
%\mbox{$\FRCVA$ } & -13,000&  -23,000 \\
%\hline
%\mbox{MtM with $\FRCVA$ - Bilateral CVA} & -133,000 & -223,000 \\
%\hline
%\end{tabular}
%\end{center}\caption{ $\FRCVA$ of the Bank - Bilateral $\CVA$  - $\Theta = 0.6$ - $\mathcal{M}$ Hypothesis - Results quoted in Euros  }
%\label{data15}
%\end{table}

\begin{table}[!h]
\begin{center}
\begin{tabular}{|c|c|c|}
\hline
$\mbox{Bank}$&$ \mbox{10 Years Swap}$ &$\mbox{30 Years Swap}$\\
%\hline  
%\mbox{MtM} & - 120,000 & - 200,000 \\
\hline  
\mbox{ $\CVA$}	& 2,500	&18,000\\
\mbox{$\FRA$ - $\mathcal{M}$ Hypothesis  }	&-10,000& -14,000 \\
\hline
\mbox{$\FRCVA$ } & -7,500&  4,000 \\
\hline
%\mbox{MtM with $\FRCVA$ - Bilateral CVA} & -127,500 &  -196,000 \\
%\hline
\end{tabular}
\end{center}\caption{ $\FRCVA$ of the Bank - $\Theta = 0.6$  - $\mathcal{M}$ Hypothesis - Results quoted in Euros  }
\label{data16}
\end{table}

%As we stressed in this paper, we conjecture that '$\mathcal{M}$ Hypothesis' is not an appropriate assumption in general,
%hence one should consider with great care any adjustment made on that basis. 
%In \pager}
%
%As a matter of comparison with the previous estimates, we represent the $\FRCVA$
%of the 30 years swap given $\Theta = 0.4$,  when $\FRA$ is set equal to  
%the funding loss mean plus a linear function of the 95 Quantile loss (denoted by $ \mbox{95Qtle}(\Phi^{1}_{\P} )$). More specifically we write:
%\beq
%\FRA^1 = < \Phi^{1} >_{\P} + x . \big( \mbox{95Qtle}(\Phi^{1}_{\P} ) - < \Phi^{1} >_{\P} \big), x \in [ 0, 1 ]. 
%\eeq
%Notice that the case $x = 0$ corresponds to the $\mathcal{M}$ Hypothesis that was evaluated previously. 
%Figure $\ref{fig:test22}$ exhibit the $\FRA$ as well as the Marked-to-Market Value of the contract as a function of the weight x. 
%Clearly, the value of the contract is very sensitive to the weight given that the 95$\%$ quantile more than offset the 
%risk-free and 'funding free' price. 
%
%\begin{figure}[!h]
%\centering
%\begin{minipage}{0.7\textwidth}
%  \centering
%  \includegraphics[width=0.7\linewidth]{30YSwapFRA4Option.png}
%  \captionof{figure}{Bank - 30 Years Swap - $\FRCVA$ and MtM after Adjustment - $\FRA$ linear function of 95 funding Qtle - Unilateral $\CVA$ - $\Theta = 0.6$}
%  \label{fig:test22}
%\end{minipage}%
%\end{figure}
%
%This naive example illustrates the difficulty at hand when assessing the $\FRA$, which we leave to further research.

\begin{figure}[!h]
\centering
\begin{minipage}{.5\textwidth}
  \centering
  \includegraphics[width=1.0\linewidth]{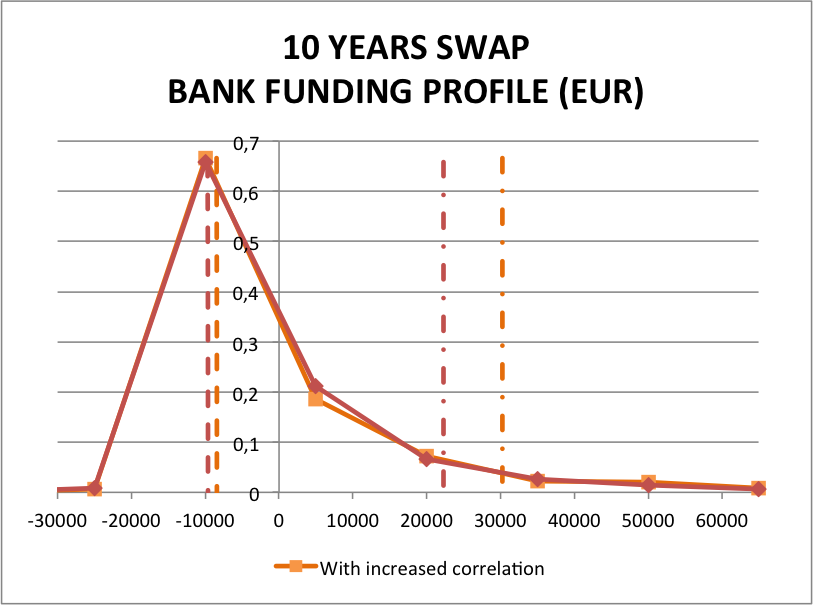}
  \captionof{figure}{Bank - 10 Years Swap\\ Funding Profile with CVA: $\Theta = 0.6$\\ and with increased correlation}
  \label{fig:test22}
\end{minipage}%
\begin{minipage}{.5\textwidth}
  \centering
  \includegraphics[width=1.0\linewidth]{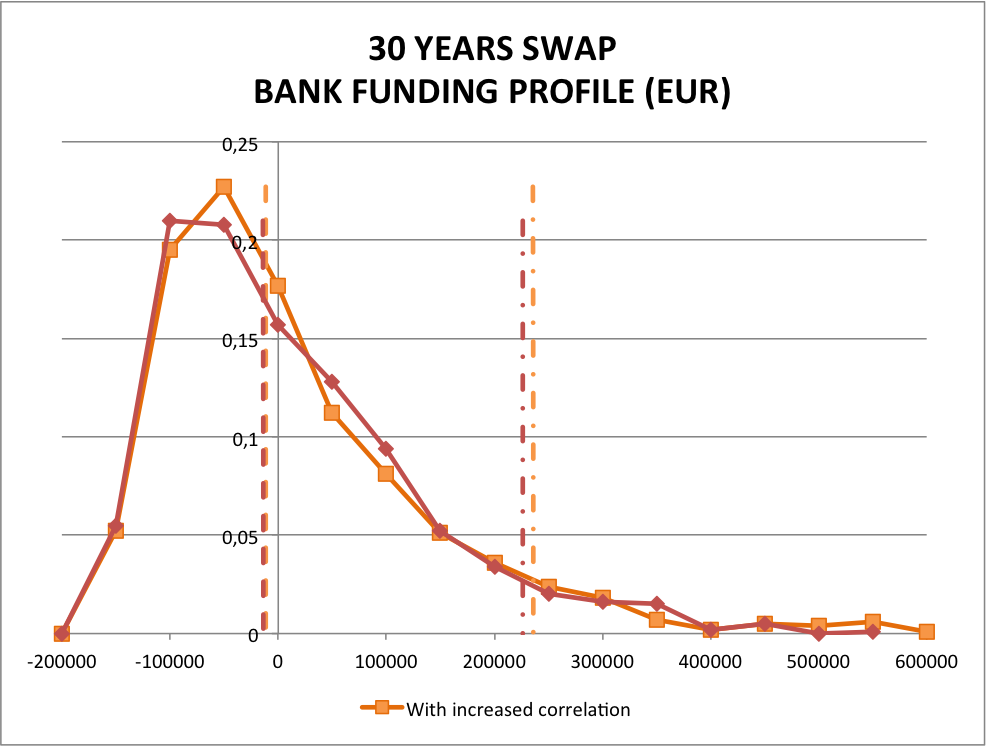}
  \captionof{figure}{ Bank - 30 Years Swap\\ Funding Profile with CVA: $\Theta = 0.6$\\ and with increased correlation}
  \label{fig:test221}
\end{minipage}
\end{figure}

Let us finally stress that our setting allows for increasing correlation among the interest rate curves and 
the credit risk spreads so as to potentially analyze right or wrong way funding risks. This is a delicate issue however and patterns are hard to discern. For example, take positive correlation between zero rates and credit spreads.  Since the funding rates $\phi$ are given by sums of interest rate and credit spreads, when correlation increases these tend to move more in the same direction and will concentrate on low or high values. 
If interest rates go down, the payer swap value goes down too, and the funding rates go down as well mostly, due to the positive correlation. The value of the contract will diminish but funding costs will diminish too, and counterparty risk will diminish as well. Some of these effects may balance and a detailed analysis is needed to decide what kind of dependence values really lead to wrong way risk. The  two graphs in Figures \ref{fig:test22} and \ref{fig:test221} are examples of the impact of increased correlation in the funding loss distribution. 

It is clear that our results are proving difficult to interpret.  We conclude
that the setting we chose may not allow for a sensible analysis of right-or-wrong way 
funding risk. We should also mention that our wrong way funding risk is similar to wrong way credit risk in the case of interest rate swaps, since it is based on correlation between interest rates and credit spreads. Clearly in one case we have the bank credit spread and in the other case we have the counterparty credit spread, but the concepts are similar.  The issue of choosing a suitable modeling approach of correlation and suitable variables for funding wrong way risk, including possibly the CDS - Bond basis, is indeed a difficult matter that needs further development. 
We will address the issue of defining a suitable framework for analysing right or wrong way funding
risk in more depth in future research.

\section{Conclusion}\label{seccon}

A key issue for
financial institutions nowadays is not to underestimate their funding risk and not being
arbitraged by competitors with a better view on the matter.
In this initial paper we argued that examining the funding loss distribution whilst computing the credit 
valuation adjustment is a sound approach from the point of view of a financial institution that cannot 
fully hedge its funding risk.
We also discussed a Weighted Cost of Funding Spread $\WCFS$ that may represent a better approach to funding cash flows modeling than the instantaneous
funding spread.
In particular, we showed examples where funding costs estimation based on the short term funding spread may
lead to underestimation of funding risk. 

In future work we need to refine our approach, investigate our setting with real market data and remove a number of simplifying assumptions that may affect our analysis, including assets traded in swapped form, assuming the funding period always reaches the final maturity with no defaults, discounting at the overnight rate,  model market risk premia properly, and study more comprehensive portfolios. We should also compare the funding loss distributions statistics with the prices for funding adjustments coming from the replication approach.

%%%%%%%%%%%%%%%%%%%%%%%%%%%%%%%%%%%%%%%%%%%%%%%%%%%%%%%%%%%%%%%%%%%%%%%%%%%%%%%%%%%%%%%%%

\end{document}